\documentclass{emulateapj}
\def\scl#1{}

\def\biggie{*}

\def\msonly#1{}

\def\comm#1           {{\tt (COMMENT: #1)}}
\def\sm#1           {{\tt (MACRO: #1)}}
\def\deg              {$^\circ$}
\def\half{{1\over2}}
\def\yr{\hbox{\,yr}}
\def\au{\hbox{\,AU}}

\slugcomment{Accepted for publication in the Astrophysical Journal}

\shorttitle{The Eccentricity Distribution of Extrasolar Planets}
\shortauthors{Juri\'{c} \& Tremaine}

\begin{document}

\title{           Dynamical Origin of Extrasolar Planet Eccentricity Distribution }

\author{
Mario Juri\'{c}\altaffilmark{1,2},
Scott Tremaine\altaffilmark{1,2}
 }

\altaffiltext{1}{Department of Astrophysical Sciences, Princeton University, Princeton, NJ
08544\label{Princeton}}
\altaffiltext{2}{School of Natural Sciences, Institute for Advanced Study, Princeton, NJ
08540\label{IAS}}


\begin{abstract}

We explore the possibility that the observed eccentricity distribution of
extrasolar planets arose through planet-planet interactions, after the
initial stage of planet formation was complete. Our results are based on $\sim
3250$ numerical integrations of ensembles of randomly constructed planetary
systems, each lasting 100 Myr. We find that for a remarkably wide range of
initial conditions the eccentricity distributions of dynamically active
planetary systems relax towards a common final equilibrium distribution, well
described by the fitting formula $dn \propto e\exp[-\half (e/0.3)^2]de$.
This distribution agrees well with the observed eccentricity distribution for
$e \gtrsim 0.2$, but predicts too few planets at lower eccentricities, even
when we exclude planets subject to tidal circularization. 
These findings suggest that a period of large-scale dynamical
instability has occurred in a significant fraction of newly formed planetary
systems, lasting 1--2 orders of magnitude longer than the $\sim 1$ Myr
interval in which gas-giant planets are assembled. This
mechanism predicts no (or weak) correlations between semimajor axis,
eccentricity, inclination, and mass in dynamically relaxed planetary
systems. An additional observational consequence of dynamical relaxation is a
significant population of planets ($\gtrsim 10$\%) that are highly inclined
($\gtrsim 25$\deg) with respect to the initial symmetry plane of the
protoplanetary disk; this
population may be detectable in transiting planets through the
Rossiter-McLaughlin effect.  

\end{abstract}

\section{                       Introduction                     }

Our understanding of the origins of planetary systems has been revolutionized in
the last decade by the discovery of over 250 extrasolar planets. These discoveries have vastly
broadened our appreciation of the diversity of possible planetary systems, and raise a number of
challenges for theories of planet formation.

One of the most important of these is the problem of large eccentricities. The
median eccentricity of the known extrasolar planets is $\simeq0.2$ (even before discarding those
planets at small semimajor axes whose orbits have likely been circularized by
tidal forces), a factor of two larger
than that of any solar-system planet except Mercury. The largest known eccentricity is 0.93, for HD 80606b
(\citealt{2001A&A...375L..27N}). The large eccentricities are a major concern because the
nearly circular, coplanar orbits of solar-system planets have been one of the strong arguments that
the planets formed from a disk since the time of Kant and Laplace. 
The mechanism by which the extrasolar planets acquired
their eccentricities, and why the eccentricities of the planets in the solar
system are so much smaller than those of the known extrasolar planets (the ``eccentricity problem'')
are still unknown. The resolution
of the eccentricity problem, and the wider question of understanding the distributions and correlations
of the dynamical and physical parameters of planets, are key milestones on the 
road towards a comprehensive theory of planet formation.

The process of planetary system formation is notoriously difficult
to study theoretically. An oversimplified but useful approach is to divide it
into two stages, based on the importance of the effects of gas and the
protoplanetary disk on the growing planets. Stage 1 lasts a few Myr, until the
dissipation of the gaseous protoplanetary disk. Irrespective of the exact
formation mechanism, at the end of this stage the gas giant planets should
have accreted their massive envelopes --- they must do so before the gas disk
disappears --- and some (very uncertain) number of smaller solid
planetesimals, planetary embryos, and planets should also be present.

Stage 2 lasts from the end of Stage 1 until the present. In Stage 2 the
evolution is driven primarily by gravitational interactions and collisions
between the planets. These processes can lead to the ejection of planets into
interstellar space or the Oort cloud, the consumption of planets by the host
star, and collisions and mergers of planets. All of these reduce the number
$N_{pl}$ of surviving planets; presumably as $N_{pl}$ declines, the system
gradually evolves to a more and more stable state, and the timescale for
future evolution is always comparable to the present age. While there is a
vast literature on the origins of solar systems that focuses on Stage 1,
investigations of Stage 2 evolution are surprisingly scarce. Yet Stage 2 is
actually easier to explore than Stage 1 in most respects, since the processes
in Stage 2 involve only the simplest possible physics: Newton's law of gravity
and Newton's laws of motion.

The first numerical explorations of Stage 2 were motivated by the suggestion
of \cite{RF96}, \cite{WM96}, and \cite{LI97} that gravitational interactions
between planets could be responsible for the large eccentricities of the
extrasolar planets\footnote{An additional motivation was the suggestion that
hot Jupiters (massive planets on circular orbits with semimajor axes
$\lesssim 0.03\au$) might be formed by tidal circularization of highly
eccentric orbits with initial semimajor axes of a few $\au$ \citep[e.g.][]{FHR01}; but, as the
sample of extrasolar planets grew, this mechanism as originally proposed proved 
unable to explain the frequency of hot Jupiters compared to planets at larger distances.
However, recent work by \cite{Nagasawa08}, as well as the frequency
of planet-star collisions observed in our simulations (Section~\ref{avgnpl}),
indicate that planet scattering in systems with multiple planets, or
in combination with the Kozai mechanism, might still plausibly explain 
the observed frequencies of hot Jupiters. This possibility deserves further investigation.}.
These numerical experiments demonstrated that planet-planet scattering could
excite eccentricities to the required levels. However, the simulations were
somewhat unrealistic in that they only lasted $\sim 10^4$--$10^6\yr$, shorter
than the likely duration of Stage 1, and typically included only a few
planets (usually two or three). Followup numerical investigations of the
evolution of unstable systems containing two identical giant planets
\citep[e.g.,][]{FHR01} found them incapable of quantitatively matching the
observed eccentricity distribution, instead producing eccentricities smaller
than those typically observed (see also the analytic arguments in 
\citealt{GS03}).  However, simulations of unstable planetary systems
containing two planets with \emph{unequal} masses \citep{Ford03}, or three or
more planets \citep{MW02,Ford03}, do appear to be capable of exciting a
sufficient fraction of planets to high eccentricities.

Other works have examined systems with more planets and over longer
timescales.  \cite{1998AJ....116.1998L} conducted a set of 16 long-term (1 to
$16$~Gyr) simulations of systems with $\sim 10$ planets, showing that stable
highly eccentric orbits are a possible outcome of long-term dynamical
evolution. \cite{2001MNRAS.325..221P} reached a similar conclusion, although
they only followed the evolution of their systems for $<10^4$ orbital periods,
which is too short to distinguish Stage 1 from Stage 2 evolution. They also
argued, following \cite{RF96}, that dynamical evolution almost inevitably leads to ejections, which could contribute to a population of
free-floating planets that may be detectable by large-scale microlensing
surveys. In the largest study to date, \cite{2003Icar..163..290A} noted the
importance of studying large ensembles of systems with similar initial
conditions, and observing the evolution of ensemble averages and
distributions.  They traced the evolution over $10^6\yr$ of four ensembles of
100 systems each, and the evolution over $10^7\yr$ of two ensembles of $50$
systems, again concluding that dynamical mechanisms are capable of producing
high eccentricities, as seen in the observations.

These studies suggest that Stage 2 evolution could play an important role in
determining the numbers, masses, and orbital elements of extrasolar planets,
but the short durations and small ensemble sizes of these integrations limit
the impact of their conclusions. They also leave many questions unanswered; in
particular, the relative importance of Stage 1 and Stage 2 in determining the
properties of planetary systems is still largely unexplored. An extreme (and
unlikely) possibility is that the dynamical properties of planets are
determined almost entirely in Stage 2, so that planetary systems are in some
kind of quasi-equilibrium state that is largely independent of the initial
conditions set at the end of Stage 1 --- just as the Maxwell-Boltzmann
distribution of gas molecules in a room is determined entirely by the
temperature and total gas mass, or the distribution of dark matter in a
galactic halo appears to be almost independent of how the halo formed
(\citealt{1997ApJ...490..493N}). An equally extreme, and almost as unlikely,
possibility is that the properties of planetary systems are determined
entirely in Stage 1, so that the distribution of planetary systems looks the
same at an age of a few Myr as it does after a few Gyr.

This paper is the beginning of a systematic investigation of the role of Stage 2 evolution in
determining the properties of planetary systems, using accurate
long-term ($10^8\yr$ or more) integrations of large (up to $N_{sys}=1000$) 
ensembles of mock planetary systems with up to 50 planets per system. In terms of the natural
metric for the computational effort required
\begin{eqnarray}
\label{eq:pp}
	PP=\hbox{(number of planets per system)} \times \\ \nonumber
	\hbox{(number of orbital periods)} \times \\ \nonumber
	\hbox{(number of systems)}\phantom{\times}
\end{eqnarray}
the results presented here have $PP = 5 \times 10^{12}$, a factor of $50$ more than any
previous exploration of Stage 2 evolution. 

In this paper we focus on the dynamical evolution of the eccentricity
distribution. In \S\ref{code} we briefly describe the integrator, the
selection of initial conditions, and simulation results. In
\S\ref{eccdist} we discuss the distributions of eccentricities of
simulated systems, their classification as ``active'', ``inactive'' and
``partially active'', and a comparison to observations. We quantify and
justify the classification criteria in
\S\ref{hillpairs}. \S\ref{inclinations} discusses the inclinations
of active systems, while \S\ref{discussion} summarizes the results and
discusses their implications.

\section{            Simulations       }
\label{code}

\subsection{The Integrator}
The principal challenge in following Stage 2 evolution is the computational intensity of the
problem. Reliable studies require (i) accurate numerical integrations of N-body systems for at least
$\sim 10^8\yr$ (our longer integrations show that relatively little evolution occurs between
$10^8$ and $10^9\yr$); (ii) the ability to follow accurately
high-eccentricity orbits and close encounters between planets (a
challenge that is not present in long solar-system integrations, where
the planets are on well-separated, low-eccentricity orbits); (iii) large ensembles of planetary systems, in order to
make statistically significant predictions and explore the wide variety of possible
outcomes.

We overcome these challenges by the combination of an integration strategy tailored for the
problem at hand, careful hand-optimization of key subroutines of the code, and the use of large
computer clusters. To efficiently simulate thousands of planetary systems in a reasonable amount of
time we have decided on a ``mix-and-match'' strategy, infrequently switching between the
Bulirsch-Stoer and hybrid symplectic integration schemes described below, depending on the conditions present in the
system being integrated. Our code makes use
of the publicly available integration routines of MERCURY6\footnote{Available at
http://www.arm.ac.uk/\~{}jec/.} (\citealt{1999MNRAS.304..793C}), and extends them to support a
high-level integration scheme as follows.

Since the initial conditions of many of the simulated planetary systems produce dynamically active
(numerous close encounters, frequent scatterings to high eccentricity orbits, etc.) and short-lived
systems, the integration of the first $10^6\yr$ is done using a high accuracy ($\epsilon =
10^{-12}$) conservative Bulirsch-Stoer integrator (``BS2'', \citealt{1999MNRAS.304..793C};
\citealt{1992nrca.book.....P}).

For the subsequent much longer integration ($10^6$--$10^8\yr$), we switch to a
hybrid-symplectic (``HYBRID'') integrator
(\citealt{1999MNRAS.304..793C}). This integrator accurately handles close
encounters between planets, without the loss of symplectic properties. It
accomplishes this by introducing a Hamiltonian splitting which, in the absence
of close encounters, is equivalent to the classical mixed-variable symplectic
scheme (\citealt{1991AJ....102.1528W}).  However, during a close encounter
(defined as two planets approaching to closer than a preset number of Hill
radii, usually 3), the HYBRID algorithm, by means of a changeover function
sensitive to mutual planet separations, redistributes the (now large)
perturbation due to the encountering planets to the Keplerian part of the
Hamiltonian. This keeps the remaining perturbation terms small, but makes the
Keplerian part analytically insoluble. The Keplerian part is therefore solved
numerically to machine precision, using a high precision Bulirsch-Stoer
integrator. The complete solution is advanced in this manner to the end of the
close encounter, when the changeover function causes the HYBRID scheme to
again become equivalent to the computationally efficient MVS. For details we
refer the reader to \cite{1999MNRAS.304..793C}.

The HYBRID scheme handles close encounters with planets accurately, but can be
susceptible to errors due to inadequate pericenter sampling in highly
eccentric orbits (\citealt{1999AJ....117.1087R}). To guard against this,
during the symplectic integration phase the conservation of total energy and
the individual orbital elements of planets in the system are continuously
monitored. If the relative energy error averaged over the past 1000 timesteps,
of size $h_{sympl}$, exceeds $10^{-4}$, or if the duration of pericenter
passage ($T_{pp}$, defined to be the time in which the true anomaly goes from
$-\pi/2$ to $+\pi/2$) of any planet become less than 5 timesteps, the code
backtracks $\Delta t = 10000$ timesteps and restarts the integration using the
high-accuracy BS2 integrator (an ``algorithm switch''). The subsequent BS2
integration phase lasts for $\Delta t = 30000h_{sympl}$, after which we return
to the hybrid-symplectic integration with $h_{sympl}$ adjusted to $1/15$th of
the smallest value of $T_{pp}$ in the system. We call this the ``HYBRID/E''
scheme since it is able to handle close pericenter passages and 
scatterings to high-eccentricity orbits, as well as close planetary encounters. 

The frequency of algorithm switches depends on the dynamical configuration of
the system being integrated. As symplecticity is violated at each algorithm
switch, it is desirable to keep these at a minimum. For the ensembles
described in the next section, 60\% of systems had no algorithm switches,
80\% had fewer than 10, while only 5\% had more than 100 in $10^8\yr$ of
integration.

We have tested the accuracy of
this scheme by comparing the results (energies, angular momenta, and time
evolution of eccentricity) of three-body integrations of highly eccentric
orbits using the BS2, HYBRID, and HYBRID/E schemes (with more steps/orbit in
the first two); typically, the errors of the HYBRID/E scheme are comparable to
those of HYBRID with $\sim 2$ times shorter timestep.
We have also tested our algorithm by comparing the results of $10^7\yr$
integrations of systems in the n10s10 ensemble (see Table~\ref{tbl.ensembles}
and the following section) with an integration done using the high-precision
Bulirsch-Stoer scheme. The final distributions of eccentricities,
inclinations, semimajor axes and masses were identical, within statistical
error. 

\begin{deluxetable\biggie}{lllllcccr}
\msonly{\rotate\tablewidth{7.7in}}
\tablecaption{Initial conditions\label{tbl.ensembles}}
\startdata
\hline
Ensemble     & $f(e)$        &  $f(I)$  &  $f(a)$          &  $f(M)$           & $N_{pl}$ &  $N_{sys}$ & Collisions & Class \\
\hline \hline
c03s00       & $0$             &  S$(I; 0.05)$ & U$(\log{}a; -1, 2)$ &  U$(\log{}M; -1, 1)$ &  3   &  500 & yes  & inactive\\
n10s10       & S$(e; 0.1)$     &  S$(I; 5.73)$ & U$(\log{}a; -1, 2)$ &  U$(\log{}M; -1, 1)$ &  10  & 1000 & no   & p. active\\
c10s10       & S$(e; 0.1)$     &  S$(I; 5.73)$ & U$(\log{}a; -1, 2)$ &  U$(\log{}M; -1, 1)$ &  10  &  200 & yes  & p. active\\
c10s00       & $0$             &  S$(I; 0.05)$ & U$(\log{}a; -1, 2)$ &  U$(\log{}M; -1, 1)$ &  10  &  150 & yes  & p. active\\
c10s30       & S$(e; 0.3)$     &  S$(I; 0.3)$  & U$(\log{}a; -1, 2)$ &  U$(\log{}M; -1, 1)$ &  10  &  200 & yes  & active\\
c50s05       & S$(e; 0.05)$    &  S$(I; 0.05)$ & U$(\log{}a; -1, 2)$ &  U$(\log{}M; -1, 1)$ &  50  &  200 & yes  & active\\
c10s40       & S$(e; 0.4)$     &  S$(I; 0.2)$  & U$(\log{}a; -1, 2)$ &  U$(\log{}M; -1, 1)$ &  10  &  500 & yes  & active\\
c10u80       & U$(e; 0, 0.8)$  &  S$(I; 3)$    & U$(\log{}a; -1, 2)$ &  U$(\log{}M; -1, 1)$ &  10  &  500 & yes  & active\enddata
\tablecomments{ Definition of initial conditions for the
ensembles of integrations. The columns (left to right) list the name of the
ensemble, the distribution functions used for drawing the initial eccentricity
$f(e)$, inclination $f(I)$ (degrees), semimajor axis $f(a)$ (in AU), mass
$f(M)$ (in units of Jupiter mass), the initial number of planets $N_{pl}$, and
the number of realizations of the ensemble $N_{sys}$. The column labeled {\em Collisions}
specifies whether planet-planet collisions were allowed to occur during the
simulation. The final column lists the ensemble classification according
to the criteria of Section~\ref{eccdist}. All systems were integrated for $10^8\yr$. $\mathrm{S}(x;
\sigma)$ is the Schwarzschild distribution (eq.~\ref{eq.sch}), $\mathrm{U}(x;
x_{min}, x_{max})$ is the uniform distribution with $x_{min} \le x <
x_{max}$.
} \end{deluxetable\biggie}

\subsection{Initial Conditions}

The selection of initial conditions would ideally be based on the predictions of Stage 1 planetary
formation theory; unfortunately, the theory is still too crude to allow this. We therefore picked
the distributions of initial conditions
(especially the eccentricities, see Figure~\ref{eccenCompInit}) for each
ensemble in a largely arbitrary fashion, with
only a minimal constraint that the planets begin in some sort of a disk.

The ensemble definitions are detailed in Table 1. We chose semimajor axes uniformly in
$\log(a)$, between $a=0.1$ and $100$~AU. Similarly, we drew the masses from a distribution uniform
in $\log(M)$, between $M = 0.1$ and $10$ Jupiter masses (this is comparable to the observed
distribution; e.g., see \citealt{2002MNRAS.335..151T}\ and \citealt{2005PThPS.158...24M}). For all but
one
ensemble we drew the eccentricity $e$ and the inclinations $I$ from a
Schwarzschild distribution\footnote{Also known as the \emph{Rayleigh distribution}.} (\citealt{1987gady.book.....B}):
\begin{equation}
\label{eq.sch}
  dN = S(x; \sigma_x) dx = \frac{x}{\sigma_x^{2}} \exp \left(-\frac{x^2}{2\sigma_x^2}\right) dx
\end{equation}
where $x$ is either $e$ or $I$, with $\sigma_e$ and $\sigma_I$ as given in Table 1. If an
eccentricity greater than 1 is drawn, the
drawing is repeated until $e < 1$ is obtained. This effective truncation of the Schwarzschild
distribution at $e = 1$ is of practical relevance in only one ensemble (c10s40, having $\sigma_e =
0.4$), as $S(x)$ falls off exponentially fast after the peak. Finally, the initial number of planets
was either 3, 10 or 50, depending on the ensemble.

The planets were approximated by homogeneous spheres of density
$\rho=1~\mathrm{g\,cm}^{-3}$. The central star was taken to have solar mass and radius. Planet-star
collisions were allowed in all ensembles. For all but one ensemble (n10s10) planet-planet collisions
were allowed as well. Both planet-star and planet-planet collisions were assumed to be fully
inelastic, resulting in momentum-conserving mergers with no fragmentation. We neglected all other
effects, such as tidal dissipation and relativistic corrections.

We have made no attempt to constrain our initial conditions to remove systems that are unstable over
short timescales, e.g., by requiring that the planets be separated by some minimum number of Hill
radii. The reason is that such systems are short-lived and hence consume a negligible fraction of
our computing resources.

For each ensemble, we constructed $N_{sys}$ realizations (planetary systems) which were then
integrated
for $10^8\yr$. This timespan corresponds to $3.2 \cdot 10^9$ and $10^5$ orbits at $a=0.1$ and
$100$~AU, respectively.

\subsection{  The average number of planets }
\label{avgnpl}

\begin{figure}
\scl{.8}
\plotone{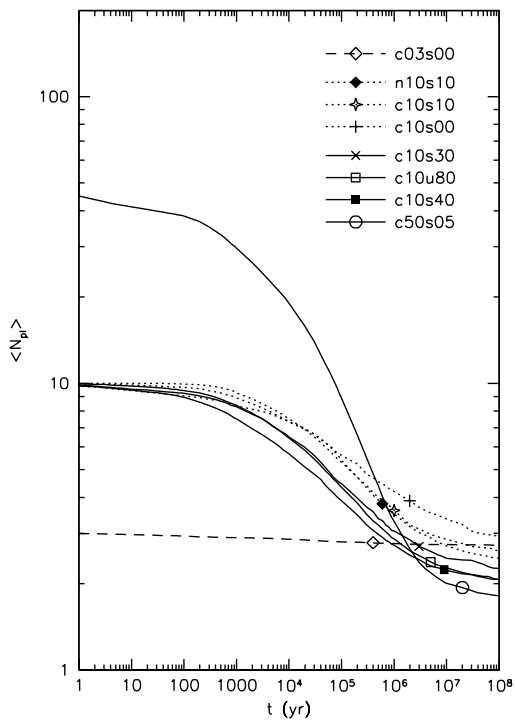}
\scl{1}
\caption{Average number of planets vs. time. Each curve shows the time
evolution of $\left<N_{pl}\right>$ for systems of a particular ensemble, as marked by different
symbols.
\label{nvst}}
\end{figure}

\begin{deluxetable}{lrrrrrrrrr}
\tablecaption{Number of planets lost to ejections, mergers, and stellar
collisions\label{tbl.decay}}
\tablewidth{0pt}
\startdata
\hline
Ensemble & \multicolumn{3}{c}{$t=10^3\yr$} & \multicolumn{3}{c}{$t=10^6\yr$} &
\multicolumn{3}{c}{$t=10^8\yr$} \\
 & E & M & S &       E & M & S &      E & M & S  \\
\hline \hline
c03s00 & 0.0 & 0.1 & 0.0 & 0.1 & 0.1 & 0.0 & 0.1 & 0.1 & 0.0 \\
n10s10 & 0.5 & 0.0 & 0.3 & 5.4 & 0.0 & 1.1 & 6.2 & 0.0 & 1.3 \\
c10s10 & 0.4 & 0.5 & 0.3 & 4.0 & 0.8 & 1.6 & 4.8 & 0.8 & 1.8 \\
c10s00 & 0.3 & 1.2 & 0.2 & 3.0 & 1.5 & 1.3 & 3.9 & 1.5 & 1.7 \\
c10s30 & 0.5 & 0.7 & 0.5 & 4.2 & 0.9 & 1.9 & 4.8 & 0.9 & 2.0 \\
c10u80 & 0.5 & 0.5 & 0.7 & 4.2 & 0.6 & 2.3 & 4.8 & 0.6 & 2.5 \\
c10s40 & 1.0 & 0.7 & 0.8 & 4.3 & 0.8 & 2.1 & 4.8 & 0.8 & 2.3 \\
c50s05 & 5.7 & 9.6 & 5.1 & 27.6 & 9.7 & 9.4 & 28.9 & 9.7 & 9.5 
\enddata
\tablecomments{
The average number of planets which by time $t$ are lost to ejections (column E),
mergers through planet-planet collisions (column M) and collisions with the star (column S). 
In ensemble n10s10 the values in column M are always $0$, because planet-planet collisions are disallowed.
}
\end{deluxetable}

The time history of the average number of planets per system is plotted in
Figure~\ref{nvst}.  The number of planets in all but one ensemble (c03s00,
which contains only three planets) exhibits a rapid dropoff, starting on a
dynamical ($\sim $--$10^3\yr$) and continuing on a secular timescale
($10^3$--$10^5\yr$). A detailed breakdown, by mode of removal of planets from
the system, is given in Table~\ref{tbl.decay}. In the first phase, the
dominant mode of removal (the dominant ``decay channel'') is through
planet-planet mergers as the randomly placed nearby planets, especially in the
inner ($a < 1$~AU) regions, collide and merge. In the first $10^3\yr$ between
3\% and 20\% of all planets are lost to planet-planet collisions. After
$10^3\yr$ there are only a few collisions, with none occurring in any of the
ensembles after $10^6\yr$.  On $t > 10^6\yr$ timescales, ejections to
interstellar space become the dominant decay channel, with between 50\% and
60\% of planets being lost in this way\footnote{c03s00 ensemble is an
exception, with an ejection fraction of 5\%.}.  A further $\sim 20$\% are lost
to collisions with the star, usually as a result of gradual eccentricity
excitation by a more massive planet. 

It is important to point out that our treatment of planets on such collision 
orbits is not complete, as we neglect the dissipative
tidal forces from the host that become significant when the pericenter is
$\lesssim 20$ stellar radii. In particular, dissipative tides may circularize the
planet orbit at small semimajor axis before it collides with the star. This
mechanism may be responsible for some, but probably not most, of the hot
Jupiters. Similarly, our treatment of ejected orbits is not complete, because
we neglect the tidal forces from the Galaxy that become significant when the
apocenter is $\gtrsim 10^3\au$. Galactic tides may cause some or even most 
planets on high-eccentricity orbits to end up in bound orbits of $\sim
10^4\au$ (analogous to the orbits in the Sun's Oort comet cloud), rather than
on unbound orbits.

In all ensembles, after approximately $10^7\yr$, the mean number of planets per system
$\langle N_{pl} \rangle$ reaches an average value between 1.8 and 3, similar to the findings of
\cite{2003Icar..163..290A} and \cite{2001MNRAS.325..221P}, despite the substantially
different initial conditions (e.g., the spherical shell of \citealt{2001MNRAS.325..221P}).
Although
$\langle N_{pl} \rangle(t)$ begins to level off at his point, indicating an end of strong dynamical
evolution it does not do so entirely. A modest power law extrapolation near $t=10^8\yr$ still
predicts a continuing decay rate of $\sim 10\%$ of planets per decade of time.

Despite this slow continuing decay, after about $2 \times 10^7\yr$ there
ceases to be any significant change in the distributions of the orbital
elements of the surviving planets (eccentricity, inclination and semimajor
axis). After $10^8\yr$, for the purposes of this paper, the simulated
ensembles can be considered to have reached an equilibrium configuration.

\section{  The eccentricity distribution }
\label{eccdist}

\begin{figure}
\scl{.55}
\plotone{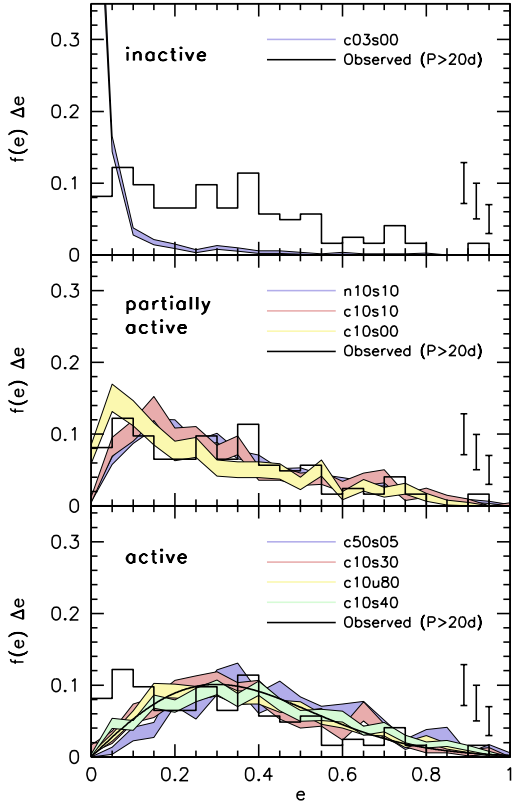}
\scl{1}
\caption{The final eccentricity distributions of the simulated ensembles. The colored bands
show the final eccentricity distributions of one ensemble classified as ``inactive'' (top), 
three classified as ``partially active'' (middle) and four ``active'' ensembles (bottom). The widths 
of the bands illustrate the $1\sigma$ Poisson uncertainty in the distribution due to the finite sizes of
simulated ensembles. The histograms show the observed distribution of eccentricities of
extrasolar planets with $P > 20$~d from \cite{Butler06}, with its typical error bars shown on the left.
The bin size is $\Delta{}e=0.05$.
Note the similarity of the final distributions of 
partially active and active ensembles to the observed eccentricity
distribution; the excess of observed planets for $e\lesssim0.2$ in the bottom panel is discussed
in \S\ref{comparisons}.
Overplotted as a smooth solid line in the bottom panel is a Schwarzschild
distribution with $\sigma_e=0.3$.
\label{eccenComp}}
\end{figure}

\begin{figure}
\plotone{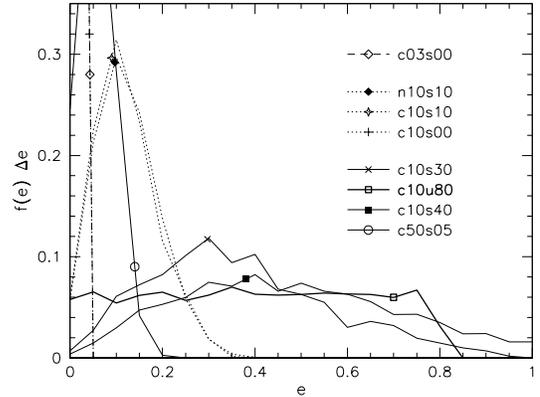}
\caption{Initial eccentricity distributions of simulated ensembles. The distributions were obtained
by binning the eccentricities of planets in the ensembles of Table~\ref{tbl.ensembles} in bins of
$\Delta{}e=0.05$. Each vertex on the plot denotes the fraction of all
planets in the eccentricity bin centered on that vertex. For clarity, the vertices are connected by
straight lines. Dashed, dotted
and solid lines correspond to the initial eccentricity distributions of ensembles whose final
distributions are shown in the top, middle and bottom panels of Figure~\ref{eccenComp}, respectively.
\label{eccenCompInit}}
\end{figure}

\subsection{Classification of outcomes}
\label{classification}

The panels of Figure~\ref{eccenComp} show the final ($t=10^8\yr$) eccentricity
distributions of the simulated ensembles (Table~\ref{tbl.ensembles}). To
visualize the expected variance due to the finite number of systems in
ensemble realizations, they are plotted as $\pm 1\sigma$ wide colored bands
(where $\sigma_i = \sqrt{N_i}/N$, $N$ is the total number of planets in the
ensemble at $t=10^8\yr$, and $N_i$ is the number of planets in the
$i^\mathrm{th}$ bin). The solid histogram overplotted on each of the panels
shows the distribution of eccentricities of 125 extrasolar planets having $P >
20$~d from \cite{Butler06}, with the period condition imposed to exclude
orbits that may have undergone tidal circularization. In
Figure~\ref{eccenCompInit}, we plot the corresponding initial eccentricity
distributions. The dashed, dotted and solid lines in
Figure~\ref{eccenCompInit} denote ensembles whose final distributions are
shown in the top, middle and bottom plots of Figure~\ref{eccenComp},
respectively.

The division into panels in Figure~\ref{eccenComp} reflects the classification
of the ensembles as either \emph{inactive} (top), \emph{partially active}
(middle), or \emph{active} (bottom). This subjective division is based on two
qualitative criteria: (i) the mutual likeness of the final eccentricity
distributions (easily seen in case of the active ensembles), and (ii) the
degree of evolution away from the initial conditions (characterized by the
fraction of planets per system surviving to $t=10^8\yr$, or the change in
shape from the initial to the final eccentricity distribution). The hope is that
this classification reflects the degree of dynamical relaxation that the
ensembles have experienced.

The first criterion clearly separates the four ``active'' ensembles from the
others (Figure~\ref{eccenComp}, bottom panel). The eccentricity distributions
of these ensembles look very similar, going to zero at $e=0$ and $e=1$ and
peaking around $e \sim 0.3$.  Quantitatively, the medians ($0.35 <
\widetilde{e} < 0.4$) as well as the widths\footnote{As measured from the semi-interquartile range, SIQR.}
($0.15 < SIQR < 0.16$) of the eccentricity distributions are all within a narrow range of values. The agreement is
particularly striking for ensembles c10u80 and c50s05, considering their
substantially different initial conditions (Figure~\ref{eccenCompInit}) ---
note in particular that in c50s05 the typical eccentricities grow, while in
c10u80 they shrink, yet both converge to a common distribution. 
The classification of these ensembles as active is also
consistent with the second criterion, as they have all undergone substantial
dynamical evolution (e.g., as evidenced by the reduction in mean number of
planets per system, Figure~\ref{nvst} and Table~\ref{tbl.decay}). All of this
indicates that the final eccentricity distribution of these systems does not
retain much memory of the initial conditions, and is primarily a result of
dynamical relaxation. Its general features are well described by the
Schwarzschild distribution of equation (\ref{eq.sch}) with $\sigma_e = 0.3$,
which is overplotted as a solid curve in the bottom panel of
Figure~\ref{eccenComp}.

The second criterion sets apart the c03s00 ensemble, the only one classified
as ``inactive''. Its initial and final eccentricity distributions are very
similar, strongly peaked at $e=0$, and inspection of the individual systems
in the ensemble shows that most have not changed significantly from their
initial state by the end of the integrations at $10^8\yr$. In the context of this paper, such ensembles are
uninteresting as Stage 2 evolution has little effect on their properties.

We assign the remaining three ensembles to an intermediate ``partially
active'' class. These ensembles \emph{have} undergone substantial evolution,
but their final eccentricity distributions are unlike those of the active
systems. The difference is most pronounced at low eccentricities ($e < 0.3$),
while the high-eccentricity ($e > 0.3$) tails are similar to those of the
active ensembles. The described behavior could be
understood as a consequence of partial relaxation. For example, it could be
that there is a range of relaxation times in different systems in a partially
active ensemble, so that some
of the systems relax fully, while the rest retain some memory
of their initial conditions. Another possibility is that the high-eccentricity
tail of the distribution settles into equilibrium rapidly, while the
equilibrium form for $e\lesssim 0.3$ is established more slowly. The partially active
distributions can also be regarded approximately as a superposition of the
distribution of fully relaxed systems, $\sim$~S$(e; 0.3)$, and the
distribution specified by the initial conditions. For our particular choice of
the initial conditions, the former dominates the high eccentricities ($e
\gtrsim 0.3$), while the latter dominates for $e \lesssim 0.3$.

In \S\ref{hillpairs}, we will return to the question of this
classification and attempt to justify it further in a more quantitative
manner.

\subsection{Comparison to observations}
\label{comparisons}

We compare the simulations with the distribution of eccentricity of all
planets in the \cite{Butler06} catalog with period $P>20$ d. This catalog
reflects strong selection effects, primarily in mass and semimajor
axis. Selection effects should not strongly affect the eccentricity
distribution since the eccentricities of the simulated planets
correlate with neither mass nor semimajor axis (this will be shown in
\S\S\ref{a}~and~\ref{m}), and since the selection effects in eccentricity
are relatively small, at least in a region ($P\lesssim 5\yr$, $e\lesssim 0.6$)
that contains most of the observed planets \citep{Cumming04}. 

Partial qualitative agreement exists between the eccentricity distributions of
the four active and three partially active ensembles and the observed
extrasolar planet distribution, shown in the bottom panel of
Figure~\ref{eccenComp}. Despite this approximate agreement, there are obvious
differences between the observed eccentricity distribution and the
distribution of active systems: (i) an excess of observed low-eccentricity
planets, and (ii) a (small) deficit of observed planets with
high eccentricities. The two are inter-related, since the distribution functions
$f(e)$ are normalized so that $\int_0^1f(e)\,de=1$. 

The excess of low-eccentricity
observed planets for $e\lesssim 0.2$ in the bottom panel of Figure
\ref{eccenComp} arises because in this region the observed systems have
$f(e)\sim\hbox{const}$ while the simulated systems have $f(e)\propto e$. Note
that the linear dependence seen in the simulated systems is a generic
consequence of dynamical relaxation: in a relaxed distribution, there is no
reason to expect a singularity in phase-space density at $e=0$ and therefore
the dependence $f(e)\propto e$ is simply due to the smaller volume of phase
space available near the origin (the radial action $J_r\propto e^2$ at fixed
semimajor axis so $dJ_r\propto e\,de$). Thus the excess of observed planets at
low eccentricities is presumably due to systems that have not
experienced the period of dynamical activity that our active-ensemble
simulations describe.

On the other hand, the simulations are highly successful in reproducing the
mid- and high-eccentricity part of the observed distribution ($e \gtrsim
0.2$). They all successfully reproduce the peak of the observed distribution
near $e \sim 0.3$, as well as the decline towards $e = 1$ and the general
shape of this decline. They appear to predict somewhat more high-eccentricity planets
than are observed, even when the $e < 0.2$ excess is taken into 
account, but the excess is only marginally statistically significant and also is
consistent with the 
effects of observational bias. In particular, \cite{Cumming04} showed that while the detection 
efficiency of radial velocity surveys is roughly constant for planets with 
$e \lesssim 0.6$, it drops rapidly beyond $e \sim 0.6$. We have conducted
simple tests, in which we apply a linearly or quadratically decreasing
detection efficiency\footnote{$p_{det} = 1 - x$ and $p_{det} = 1 - x^2$, where
  $x = (e-0.6)/0.4$.} to simulated data at $e > 0.6$, and find that the correction works in 
the direction of improving the agreement with our simulations. But overall,
its effect on the shape of the observed distribution is negligible for the
current data.

\begin{figure}
\scl{1}
\plotone{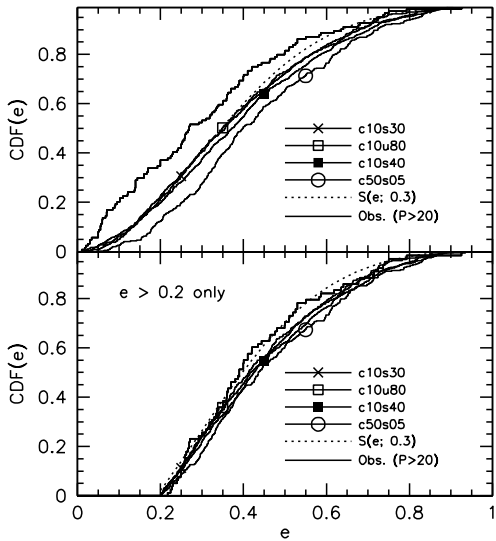}
\scl{1}
\caption{ The cumulative distribution functions of eccentricity in the
observed sample (solid line with no symbol) and the four active ensembles
(solid lines with symbols), and the cumulative distribution function of the
Schwarzschild distribution S$(e; 0.3)$ (eq.\ \ref{eq.sch}; dotted line) for
all planets (top) and planets with $e > 0.2$ (bottom).
\label{fig.kstest}}
\end{figure}

\begin{deluxetable}{lccccc}
\tablewidth{0pt}
\tablecaption{KS tests of eccentricity distributions (active ensembles)\label{tbl.kstest}}
\startdata

 All      & c10u80 & c10s40 & c50s05 & Observed & S$(e; 0.3)$  \\
 \hline\hline
 c10s30 & $0.99$ & $0.42$ & $9 \times 10^{-4}$ & $5 \times 10^{-3}$ & $0.41$ \\
 c10u80 & $\cdot$ & $0.21$ & $3 \times 10^{-5}$ & $2 \times 10^{-3}$ & $0.15$ \\
 c10s40 & $\cdot$ & $\cdot$ & $3 \times 10^{-3}$ & $5 \times 10^{-4}$ & $8 \times 10^{-4}$ \\
 c50s05 & $\cdot$ & $\cdot$ & $\cdot$ & $5 \times 10^{-7}$ & $2 \times 10^{-5}$ \\
 Observed & $\cdot$ & $\cdot$ & $\cdot$ & $\cdot$ & $2 \times 10^{-4}$ \\
 \hline

 & & & & &  \\
 $e > 0.2$  & c10u80 & c10s40 & c50s05 & Observed & S$(e; 0.3)$  \\
 \hline\hline
 c10s30 & $0.99$ & $0.57$ & $0.28$ & $0.54$ & $0.21$ \\
 c10u80 & $\cdot$ & $0.57$ & $0.14$ & $0.30$ & $0.02$ \\
 c10s40 & $\cdot$ & $\cdot$ & $0.48$ & $0.13$ & $ 1 \times 10^{-4}$ \\
 c50s05 & $\cdot$ & $\cdot$ & $\cdot$ & $0.07$ & $ 2 \times 10^{-4}$ \\
 Observed & $\cdot$ & $\cdot$ & $\cdot$ & $\cdot$ & $0.53$
\enddata
\tablecomments{ The $p$-values of Kolmogorov-Smirnov tests between 
the eccentricity distributions of active ensembles (c10s30, c10u80, c10s40
and c50s05), the observed sample of extrasolar planets with $P>20$ d
(``Observed''), and the Schwarzschild distribution with $\sigma_e=0.3$ (last
column).  When testing against the S$(e; 0.3)$ distribution the null
hypothesis is that the sample was drawn from it (one-sample KS test).  Everywhere
else, the null hypothesis is that the pair of samples being tested was drawn
from the same (but unspecified) underlying distribution (two-sample KS test).
The top set of rows shows the results when planets of all eccentricities are
included in the test.  The bottom set shows the results for the subsample with
$e > 0.2$.
}
\end{deluxetable}

\begin{deluxetable}{lccccc}
\tablewidth{0pt}
\tablecaption{KS tests of eccentricity distributions (partially active ensembles)\label{tbl.kstest.pa}}
\startdata

 All  & c10s10 & c10s00 & Observed & S$(e; 0.3)$  \\
 \hline\hline
 n10s10 & $0.11$ & $0$ & $0.15$ & $0$ \\
 c10s10 & $\cdot$ & $7 \times 10^{-6}$ & $0.32$ & $0$ \\
 c10s00 & $\cdot$ & $\cdot$ & $0.01$ & $0$ \\
 Observed & $\cdot$ & $\cdot$ & $\cdot$ & $$ \\
 \hline

 & & & & &  \\
 $e > 0.2$ & c10s10 & c10s00 & Observed & S$(e; 0.3)$  \\
 \hline\hline
 n10s10 & $0.75$ & $0.55$ & $0.72$ & $1 \times 10^{-3}$ \\
 c10s10 & $\cdot$ & $0.55$ & $0.80$ & $0.06$ \\
 c10s00 & $\cdot$ & $\cdot$ & $0.68$ & $0.01$ \\
 Observed & $\cdot$ & $\cdot$ & $\cdot$ & $0.53$

\enddata
\tablecomments{
Analogous to Table~\ref{tbl.kstest}, but for partially active ensembles.
}
\end{deluxetable}

\msonly{\clearpage}

We quantify the discussed similarities and differences using
Figure~\ref{fig.kstest} and Tables~\ref{tbl.kstest}~and~\ref{tbl.kstest.pa}.
Figure~\ref{fig.kstest} compares the final cumulative eccentricity
distributions of the active ensembles, the observed distribution, and the
Schwarzschild distribution S$(e; \sigma_e = 0.3)$ which captures the general
features of the simulated distributions. The comparison is
made both over the entire range of eccentricities (top panel), and restricted
to the range $e > 0.2$ to exclude possible contamination by inactive systems in the observed sample
(bottom panel).  Table~\ref{tbl.kstest} shows the $p$-values of the
Kolmogorov-Smirnov\footnote{The same tests were repeated using Pearson's
$\chi^2$ statistic and led to qualitatively similar conclusions. However, for
these data sets the $\chi^2$ statistic has less distinguishing power than the KS statistic
(typical $p$-values were a factor of two larger).} statistic calculated
between pairs of all active ensembles, the observed distribution, and the
S$(e; \sigma_e = 0.3)$ distribution. Table~\ref{tbl.kstest.pa} shows the same
statistic for partially active ensembles.

At a 5\% significance level\footnote{
In this paper we adopt the $\alpha=0.05$ significance level as the threshold for 
rejecting the null hypothesis that two samples were drawn from the same
distribution. That is, if at least 5\% of pairs of samples 
randomly drawn from a given distribution would differ by more than the observed amount
as measured by their KS statistic, we conclude that the hypothesis cannot be rejected.
}
the eccentricity distributions of three of the four active ensembles
(c10s30, c10u80, c10s40) are pairwise\footnote{We use the term ``pairwise'' to
emphasize that these are two-sample tests, and that they do not test whether
\emph{all} ensembles are \emph{simultaneously} consistent with being drawn
from the same underlying distribution. For example, three ensembles A,
B and C with two-sample KS statistic $p$-values $p(AB)$, $p(BC)$, and $p(AC)$
above a significance level $\alpha$ may produce $p(ABC) < \alpha$ in a
three-sample test.}  consistent with being drawn from the same eccentricity
distribution (Table~\ref{tbl.kstest}), despite their quite different initial
conditions (Table~\ref{tbl.ensembles}). The smallest $p$-values are obtained
in tests involving c50s05, which began with 50 planets per system, compared to
10 for the other active ensembles. This difference may point to a dependence
of the finer details of the final eccentricity distribution on 
the initial conditions.  When restricted to the subsample with $e>0.2$, the
distributions of \emph{all} active ensembles are pairwise consistent. Analogous tests
of the partially active ensembles (Table~\ref{tbl.kstest.pa}) reveal similar
results: substantial differences in the eccentricity distributions of the full
samples but consistency among the subsamples with $e>0.2$. 

Using the same statistic, we compare the final $e > 0.2$ distributions of
active and partially active ensembles with the observations
(Tables~\ref{tbl.kstest}~and~\ref{tbl.kstest.pa}, the columns and rows labeled
``Observed''). In all cases, the simulated $e > 0.2$ distributions and the
observed distribution are consistent with being drawn from the same underlying
distribution at the 5\% significance level. We interpret this as evidence that
the high-eccentricity component ($e>0.2$) of active and partially active
ensembles is populated by planets from dynamically relaxed systems, as is the
majority of the high-eccentricity component of the observed extrasolar planet
population\footnote{Other mechanisms, such as Kozai oscillations due to a
companion\citep{wm03,2007ApJ...669.1298F}, may be responsible for a minority of
high-eccentricity planets.}.  Taking into account the simplifying
approximations of our simulation (giant planets only, no debris, gas disk or
any other influences, no binary companions), the exclusion of all
non-gravitational effects (e.g., tidal effects), and the likely differences
between our assumed initial distribution of masses and orbital elements and
the actual distribution, the agreement obtained for $e > 0.2$ is quite
remarkable.

The Schwarzschild distribution S$(e;\sigma_e=0.3)$ provides an approximate qualitative 
representation of the eccentricity distribution of the active ensembles
(Figure \ref{eccenComp}, bottom panel) but is not quantitatively consistent
with several of them according to the KS statistic---probably we could improve
the consistency by fitting $\sigma_e$ to the eccentricity distributions, but
this would provide only an illusion of greater accuracy.

In the scheme introduced in
\S\ref{classification}, the ensemble of observed planets would be
classified as partially active. Its distribution of eccentricities may be
decomposed into two components. One, resulting from dynamical relaxation,
dominates the $e > 0.2$ regime, contains 75\% of $P > 20$~d (or 55\% of all) planets, 
and agrees well with the $e > 0.2$ distributions of the active and partially active ensembles
in our simulations. The other contains 25\% of $P > 20$~d (or 45\% of all) planets and dominates the
$e < 0.2$ regime; here the eccentricities were set by other processes,
possibly the (unknown) initial conditions or damping by low-mass planets,
planetesimals, or residual gas. With this decomposition in mind,
for the remainder of this paper we will mostly concentrate on the properties
of the relaxed component (the active ensembles).

\subsection{  Time evolution }

\begin{figure}
\scl{.6}
\plotone{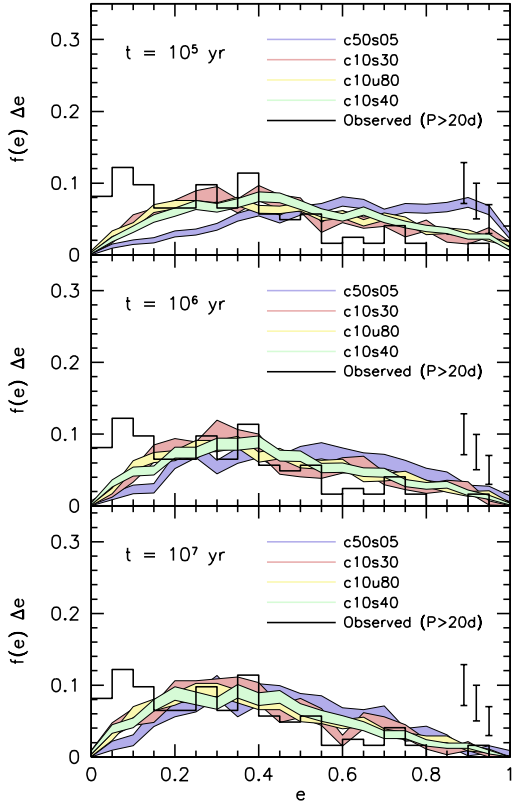}
\scl{1}
\caption{ Time evolution of eccentricity distributions of active
ensembles. The top, middle, and bottom panel show the eccentricity
distributions at $t=10^5$, $10^6$ and $10^7\yr$ respectively.  The meaning of
symbols is the same as in Figure~\ref{eccenComp}, where the corresponding
distributions at $10^8\yr$ are shown in the bottom panel.
\label{eccevol}}
\end{figure}

The panels of Figure~\ref{eccevol} show the time evolution of the eccentricity
distributions of active ensembles from $t=10^5\yr$ (top) to $t=10^7\yr$
(bottom). At $t=10^5\yr$ the distributions are still dissimilar,
especially that of the c50s05 ensemble which is still undergoing strong
dynamical activity (a consequence of the larger initial number of planets).
By $t=10^6\yr$ the fraction of high eccentricity planets has
been reduced in all ensembles, with a simultaneous increase in the frequency 
of planets of moderate eccentricity. At $t=10^7\yr$ the eccentricity distributions have
largely converged to a common characteristic shape, with the biggest change
from $10^7$ to $10^8\yr$ being a further reduction in the number of high
eccentricity planets, primarily by ejections (see
bottom panel of Figure \ref{eccenComp}).

\subsection{  Influence of collisions }

\begin{figure}[!ht]
\plotone{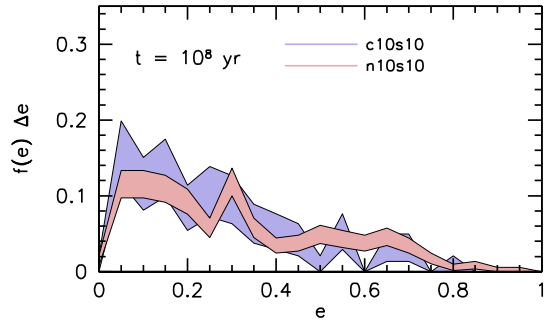}
\caption{Comparison of eccentricity distributions of planets having $a < 1$~AU at $t=10^8\yr$, in ensembles n10s10 and c10s10.
The meaning of symbols is the same as in Figure~\ref{eccenComp}.
\label{cVSn}}
\end{figure}

By comparing ensembles n10s10 and c10s10 we tested the influence of
planet-planet collisions on the shape of the final eccentricity
distribution. These two ensembles share the same initial conditions, except
that in the case of c10s10 collisions were allowed to occur, while for n10s10 they
were not (i.e., the planets were assumed to have density
$\rho=1~\mathrm{g\,cm}^{-3}$ in the first case, and to be point masses in the
second). We found no significant difference in the outcomes of these two cases
(Figure~\ref{eccenComp}, middle panel), and the final eccentricity
distributions are consistent with being drawn from the same distribution at
the 10\% significance level (Table~\ref{tbl.kstest.pa}). Since the effects of
planet-planet collisions are likely to be most noticeable at small semimajor
axes, we also compared the distributions of n10s10 and c10s10 planets
having semimajor axis $a < 1\au$ (Figure~\ref{cVSn}), but again found no significant difference.

This outcome was not unexpected, since we have already observed that the final
eccentricity distribution is established over long timescales ($\sim
10^5$--$10^8\yr$, Figure~\ref{eccevol}), while planet-planet collisions are
infrequent events (Table \ref{tbl.decay}) which preferentially occur early ($t
\lesssim 10^4\yr$). Physically, the ratio of the escape speed from the planets
to the typical encounter speeds between planets at $1\au$ is large enough that
the planets act like point masses. 

\subsection{  Dependence on semimajor axis }
\label{a}

\begin{figure\biggie}
\plotone{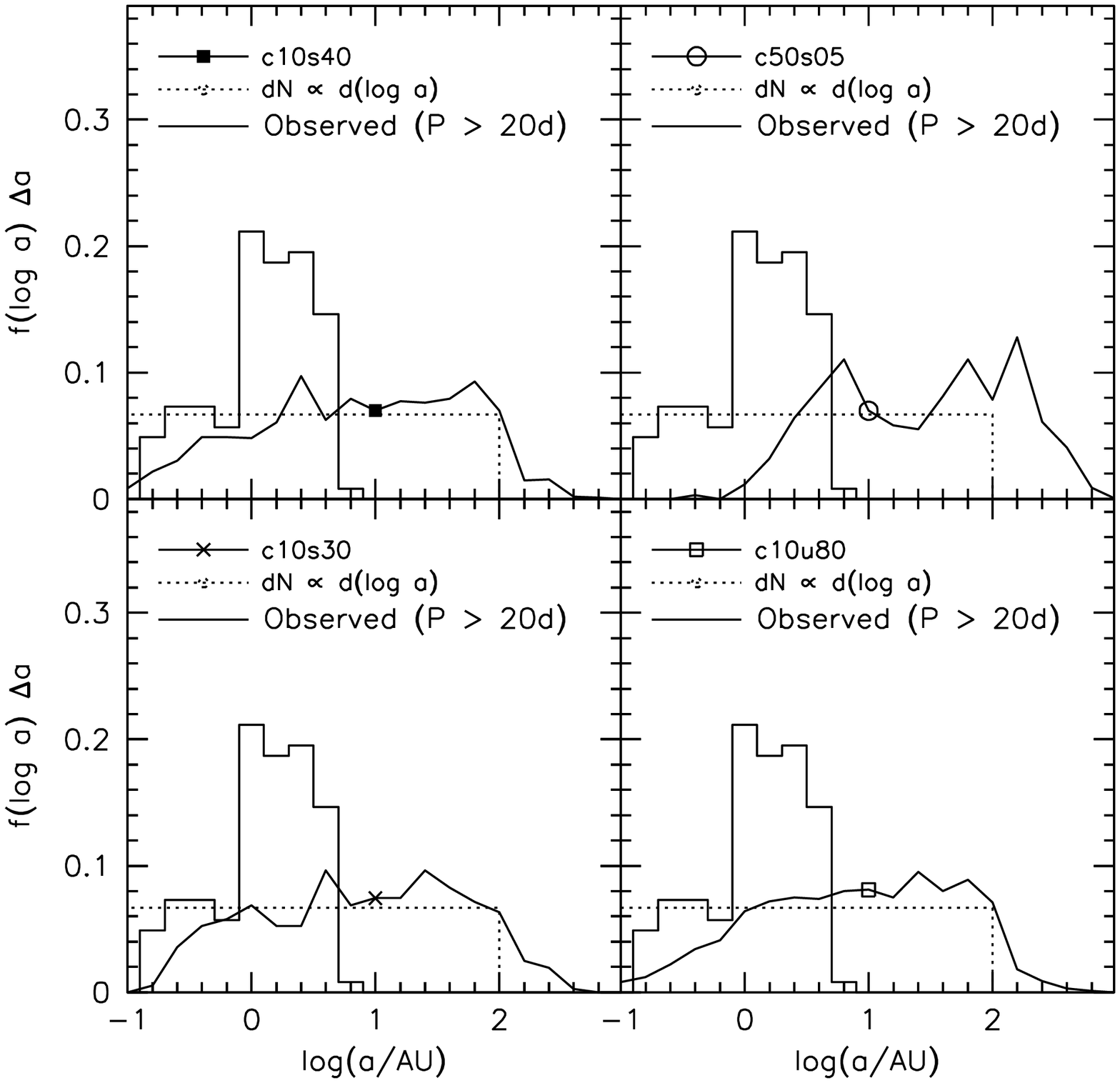}
\caption{The distribution of semimajor axes in the four active ensembles. The
semimajor axes of known exoplanets having $P > 20$~d are plotted
as a histogram. Note that the two distributions are not directly comparable, due to the
significant observational biases existing in the latter. The bin size is
$\Delta{}\log{a}=0.2$.
\label{lahist}}
\end{figure\biggie}

\begin{figure}
\scl{.55}
\plotone{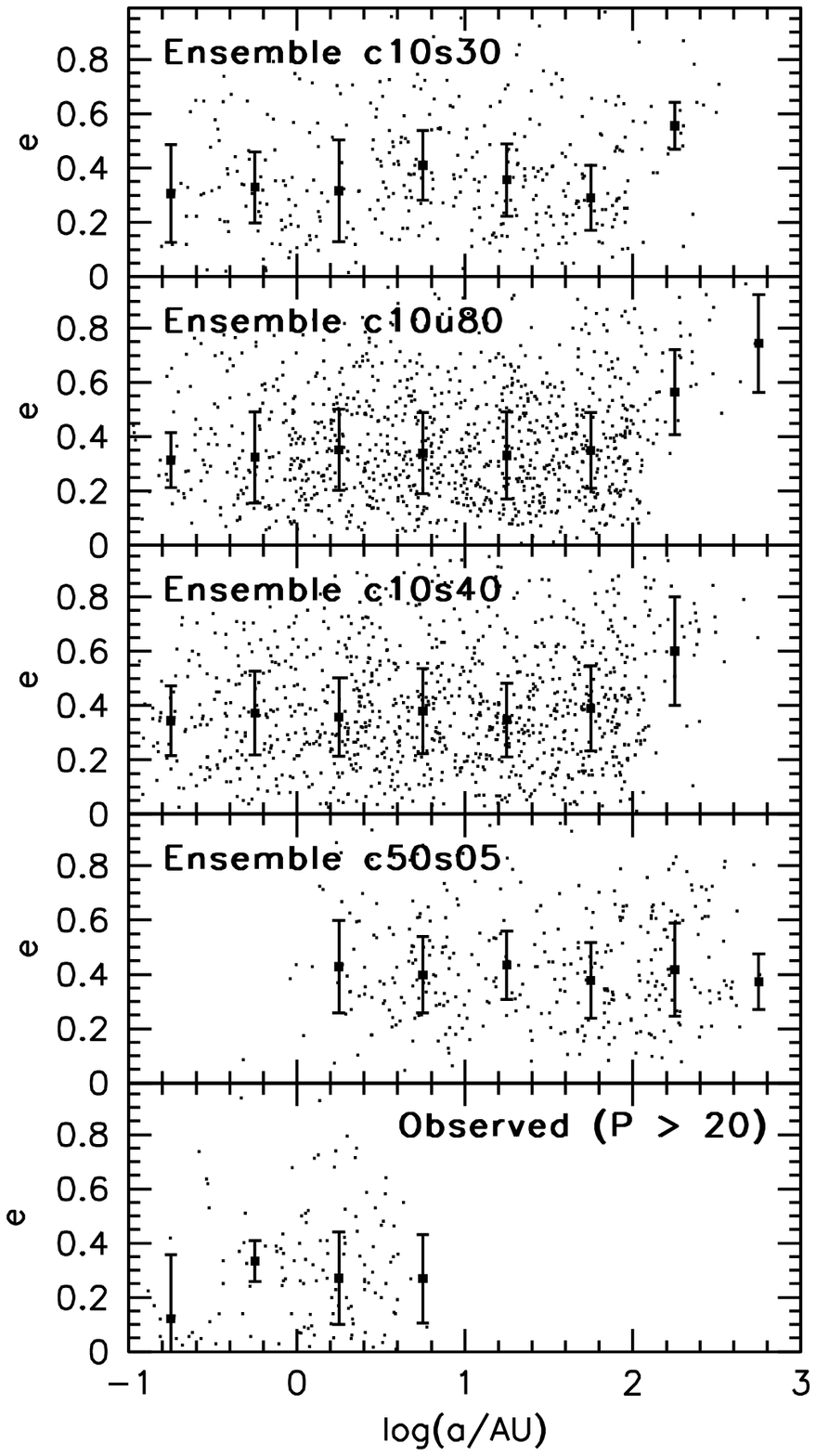}
\scl{1}
\caption{The correlation of semimajor axis and eccentricity of active
  ensembles (top four panels) and the observed ($P > 20$~d) sample (bottom
  panel). The error bars show the semi-interquartile range.
\label{logaeplotComp}}
\end{figure}

Figures~\ref{lahist}~and~\ref{logaeplotComp} examine the semimajor axis
distribution and the correlation of eccentricity with semimajor axis in the four
active ensembles.

The initial semimajor axis distribution of all of our ensembles was uniform in
$\log a$. The final distribution in the active ensembles shows depletion at
low $a$ and some spreading beyond $100\au$ (the initial outer limit). Neither
is particularly surprising given that interactions are strongest at small
semimajor axes (thus the depletion at low $a$), and that planet-planet
scattering tends to spread out the semimajor axis distribution. The
efficiency of depletion is particularly striking in case of c50s05, where
fewer than 2\% of planets remain on orbits closer than $a = 1\au$;
nevertheless, in general it appears that the semimajor axis distribution in
active systems retains a strong memory of the initial conditions.

Of greater consequence for comparisons of the eccentricity distributions with
the observations is the correlation of eccentricity and semimajor axis. We
find no significant $e$-$a$ correlation, except for a small subset (less than
5\% of the total) of planets scattered to high-eccentricity orbits at $a >
100$~AU (Figure~\ref{logaeplotComp}); this correlation is expected, since such
orbits are likely to result from close encounters with planets at smaller
radii, and hence must have pericenters $q\lesssim 100\au$ so $a=q/(1-e)$ is
correlated with $e$. The widths of the eccentricity distributions at fixed $a$
remain approximately constant 
(semi-interquartile range SIQR$\simeq 0.15$) over the entire range of
semimajor axes, indicating that the shape of the eccentricity distribution
does not appreciably vary with $a$ either. This has already been implicitly
assumed in \S\ref{comparisons} where we compared the distributions of the
entire simulated sample (with median $\widetilde{a}\sim 7.5 \au$--$8.5\au$ in c10-
ensembles and $\widetilde{a}\sim 34\au$ in c50s05) to
the observed sample ($\widetilde{a} \sim 1.3\au$).

\subsection{  Mass--eccentricity correlations }
\label{m}

\begin{figure\biggie}
\plotone{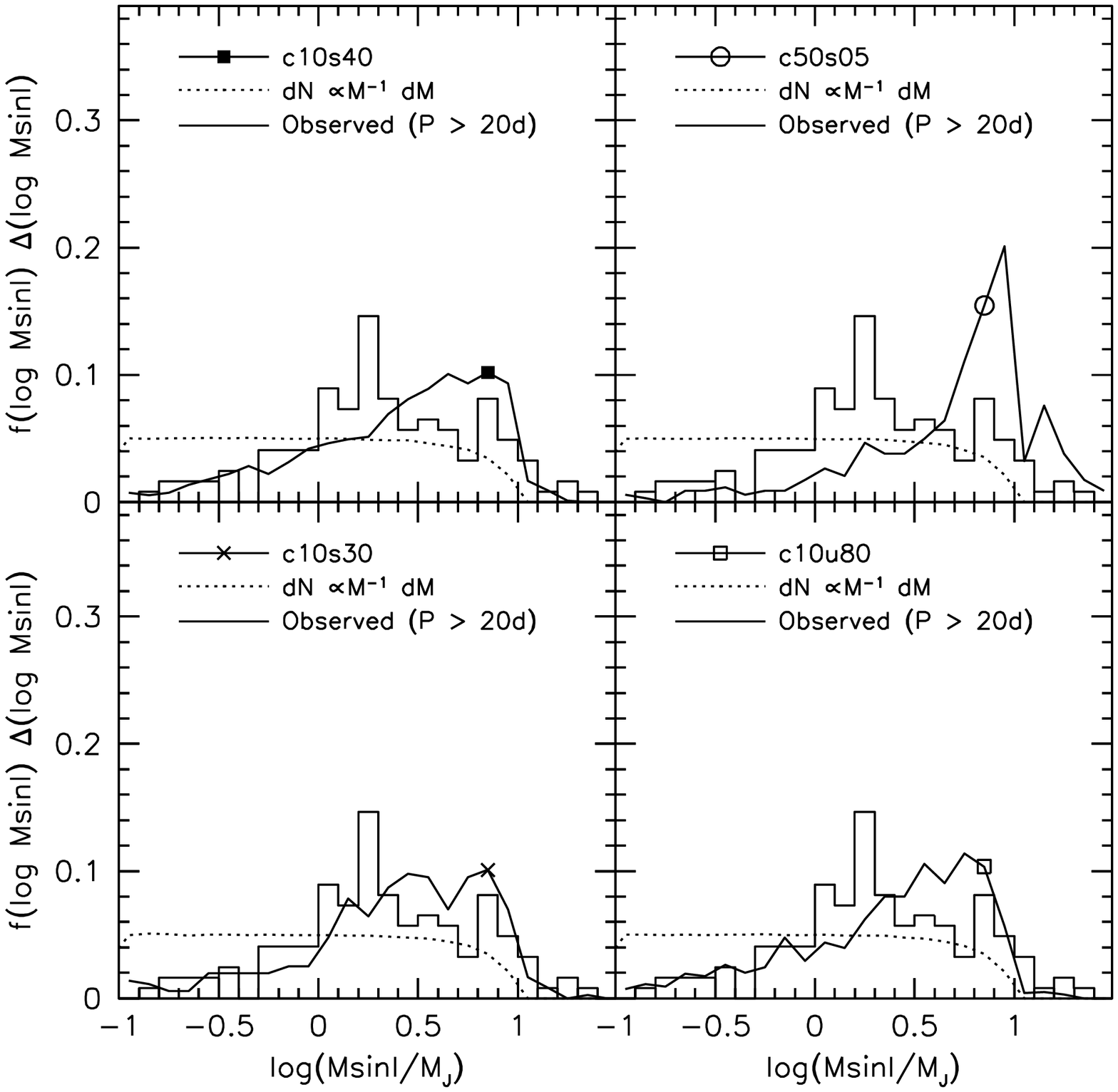}
\caption{ Distribution of $M \sin I$ at $t = 10^8\yr$ in the four active
ensembles (solid line), compared to the observed distribution of $M \sin I$
(solid histogram), and the distribution of initial conditions ($dN \propto
d\log{M}$, $-1 \leq \log(M/M_J) \leq 1$; dotted line).  To obtain $M \sin I$
in the simulations we assume that the orbit normals are uniformly distributed
on a sphere. The bin size is $\Delta{}\log(M\sin I)=0.1$.
\label{lmhist}}
\end{figure\biggie}

\begin{figure}
\scl{.55}
\plotone{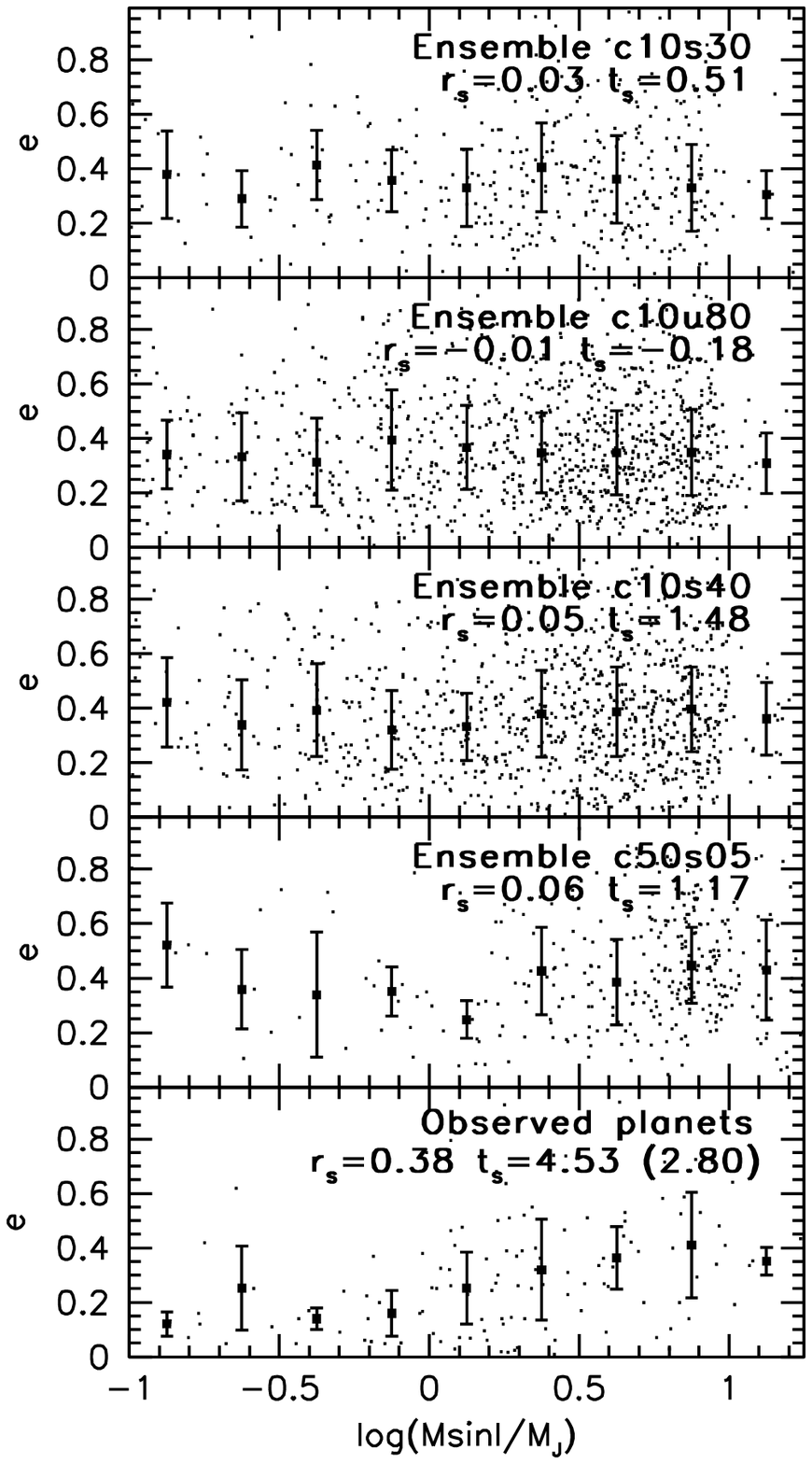}
\scl{1}
\caption{ Dependence of eccentricity on $M \sin I$ at $10^8\yr$ for the four
active ensembles (top four panels) and the observed ($P > 20$~d)
planets. Dots mark the eccentricity and $M \sin I $ of each planet. To obtain
$M \sin I$ in the simulations we assume that the orbit normals are
uniformly distributed on a sphere. The median and semi-interquartile ranges of
$e$, calculated in bins of size $\Delta\log(M \sin I) = 0.25$, are marked by the
solid squares and error bars, respectively.
The values of the Spearman rank correlation coefficient $r_s$ evaluated over all
planets in the sample and the corresponding normally distributed
$N(\mu=0,\sigma=1)$ variable $t_s$ are given in the upper right corner of each
panel. The number in parentheses in the bottom panel is the value of $t_s$ for
a subsample satisfying $M\sin I > M_J$.
\label{eccmcor}}
\end{figure}

In Figure~\ref{lmhist} we show the final ($t=10^8\yr$) distributions of $M
\sin I$ in the four active ensembles (solid lines), compared to the observed
$M \sin I$ distribution (solid histograms) and the initial mass distribution
function ($dN \propto M^{-1} dM$, dotted line). To obtain $M \sin I$ from
simulations, we assume that the orbit normals are uniformly and randomly
distributed on a sphere and assign the inclinations accordingly. Note that the
observed distribution of exoplanet $M\sin I$ is heavily biased by the
difficulty of detecting low-mass planets; the true distribution almost
certainly is steeper and extends to lower values than the measured one. 

The mass distributions for the three ensembles that began with $N_{pl}=10$
planets per system (c10s40, c10s30, and c10u80) converge to a similar final
shape by $t=10^8\yr$. The $M \sin I$ distribution of c50s05 (started with
$N_{pl}=50$ planets per system) is different: strongly peaked around $M \sin I
\sim 7.5 M_J$, with an extended tail beyond $M \sin I = 10 M_J$ due to a
significantly higher fraction of mergers than the other ensembles
(Table~\ref{tbl.decay}).

Relative to the initial conditions (dotted line), all of these simulations
show a sharp reduction in the fraction of planets with small masses
($M\lesssim M_J$), arising because low-mass planets are more readily removed
from the systems, and mergers shift the distribution towards higher
masses. Compared to the observed distribution, these three ensembles show a
statistically significant deficit of planets in the range 
$M_J \lesssim M\sin I \lesssim 3 M_J$ but agree with the observed distribution for
$M\sin I \gtrsim 3M_J$ (KS test; 5\% significance level). The difference at low masses is not due to 
selection effects in the observed sample, since these preferentially 
delete the lower mass planets, not the higher mass ones. Simulations using 
a steeper initial mass distribution may result in a better fit to the observational data.

\begin{figure\biggie}
\plotone{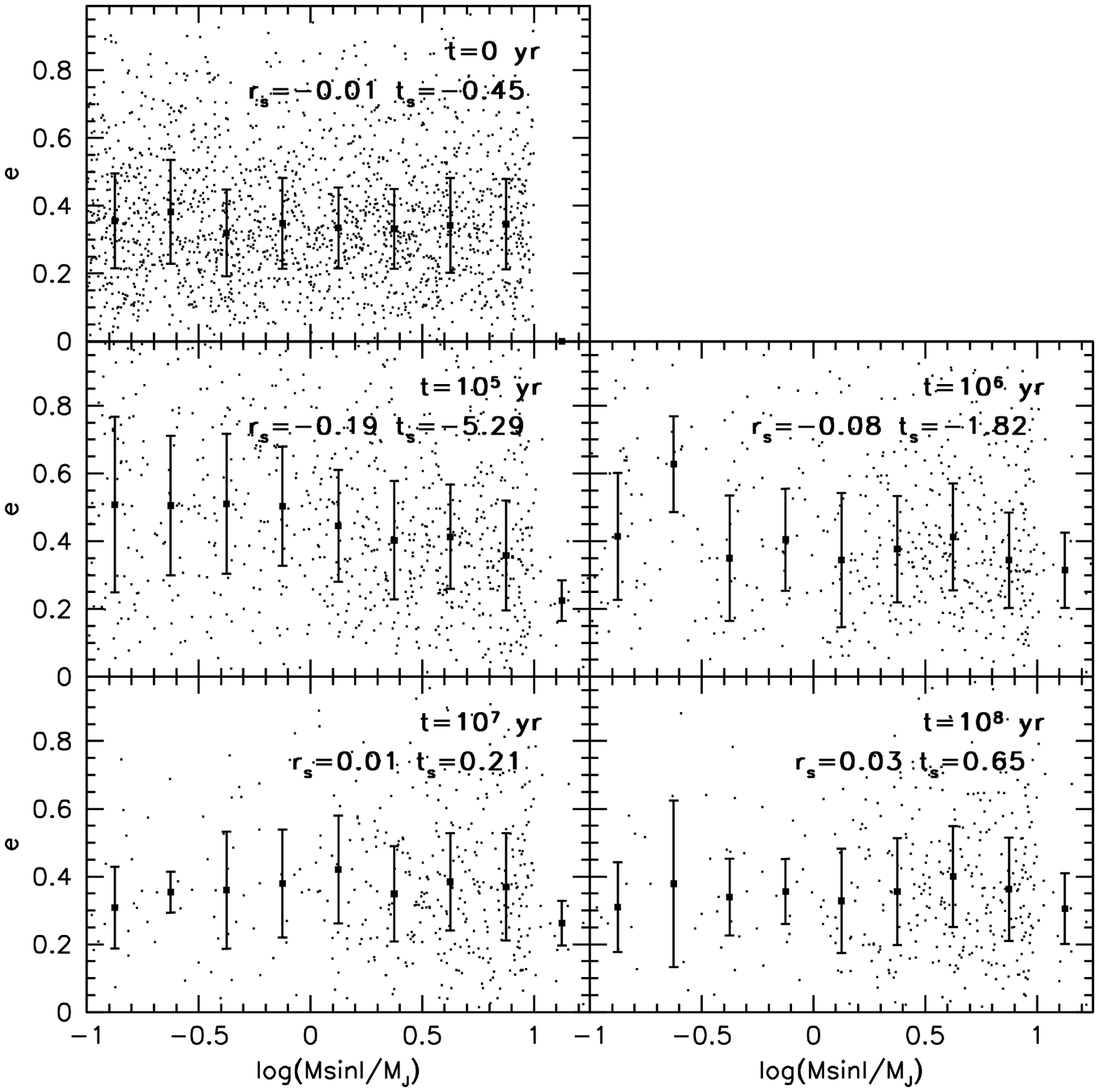}
\caption{ The evolution of the mass-eccentricity correlation in the c10s30 ensemble. The panels (left to
right, top to bottom) show the masses and eccentricities of planets
at times $t=0$, $10^5$, $10^6$, $10^7$, and $10^8\yr$. The
meaning of the symbols is the same as in Figure~\ref{eccmcor}.
\label{eccmevol}}
\end{figure\biggie}

Figure~\ref{eccmcor} shows the correlation of eccentricity and $M \sin I$ at
$t=10^8\yr$ in the four active ensembles and in the observed sample. We find
non-zero but statistically insignificant positive mass-eccentricity correlations
in the simulated ensembles.  The median eccentricity of planets in different
mass bins is roughly constant ($\widetilde{e} \sim 0.35$), as is the width
$e_{\mathrm{SIQR}} \sim 0.15$ of the eccentricity distributions as measured by
the semi-interquartile range. The lack of strong correlation between $e$ and
$M \sin I$ may be surprising given that one might expect some kind of
equipartition, in which the most massive planets acquire the smallest
eccentricities in planet-planet scattering. Indeed, as we show in
Figure~\ref{eccmevol}, such correlations are present at some times during the
simulation, although they are never as strong as equipartition of radial
energies would predict.

In the panels of Figure~\ref{eccmevol} we show the evolution of the
mass-eccentricity correlation for one ensemble (the other active ensembles
exhibit similar behavior). The top left panel shows the dependence of average
eccentricity on mass at $t=0$, and the subsequent panels (left to right, top
to bottom) show the $e$ vs.\ $M\sin I$ correlation at $t=10^5$, $10^6$, $10^7$,
and $10^8\yr$. The correlation of mass and eccentricity changes during the
integration. Initially ($0 < t < 10^5\yr$), the median
eccentricity of low-mass planets grows, with the median eccentricity of high-mass
planets also growing but by a smaller amount. This is followed by a period of
decline of the median eccentricity of both low- and high-mass planets ($10^5 < t <
10^7\yr$ in Figure~\ref{eccmevol}), with eccentricities evolving to a
mass-independent median value $\widetilde{e}\sim 0.35$ at $t > 10^7\yr$.

This behavior is a consequence of dynamical interactions in the system. At
early times ($0 < t < 10^5\yr$), the low-mass planets are easily excited to
higher eccentricity orbits by their massive counterparts. As these
high-eccentricity planets are gradually removed from the system (through close
encounters, ejections, or collisions with the star) the average eccentricity
of the remaining planets decreases. This is supported by the finding that
$\sim$80\% of planets that are excited to $e > 0.6$ are removed from the
system by $10^8\yr$.  The median eccentricity of the high-mass planets initially
increases, and then gradually decreases over time through the removal of
planets on high-eccentricity orbits and the damping of eccentricity by lower
mass planets. These processes continue until enough planets are removed from
the system and the orbits of the remaining planets become sufficiently
separated, thus rendering the system stable for the remainder of the
integration.

The observed sample (Figure~\ref{eccmcor}, bottom panel) shows a positive
correlation between mass and eccentricity at the $\sim 5.5 \sigma$ level. This
signal comes largely from a difference in median eccentricity between the
planets with $M\sin I \lesssim M_J$ and those with $M\sin I\gtrsim 3M_J$.  If
the sample is restricted to the subset with $M\sin I > M_J$, its significance drops
to $\sim 2.8\sigma$.  Some of this correlation may result from selection
effects (planets are harder to detect either if the mass is low or the
eccentricity is high), so we prefer to wait for more data before investigating
the implications of a possible mass-eccentricity correlation. 

\section{  Inclinations of active systems }
\label{inclinations}

\begin{figure\biggie}[!ht]
\plotone{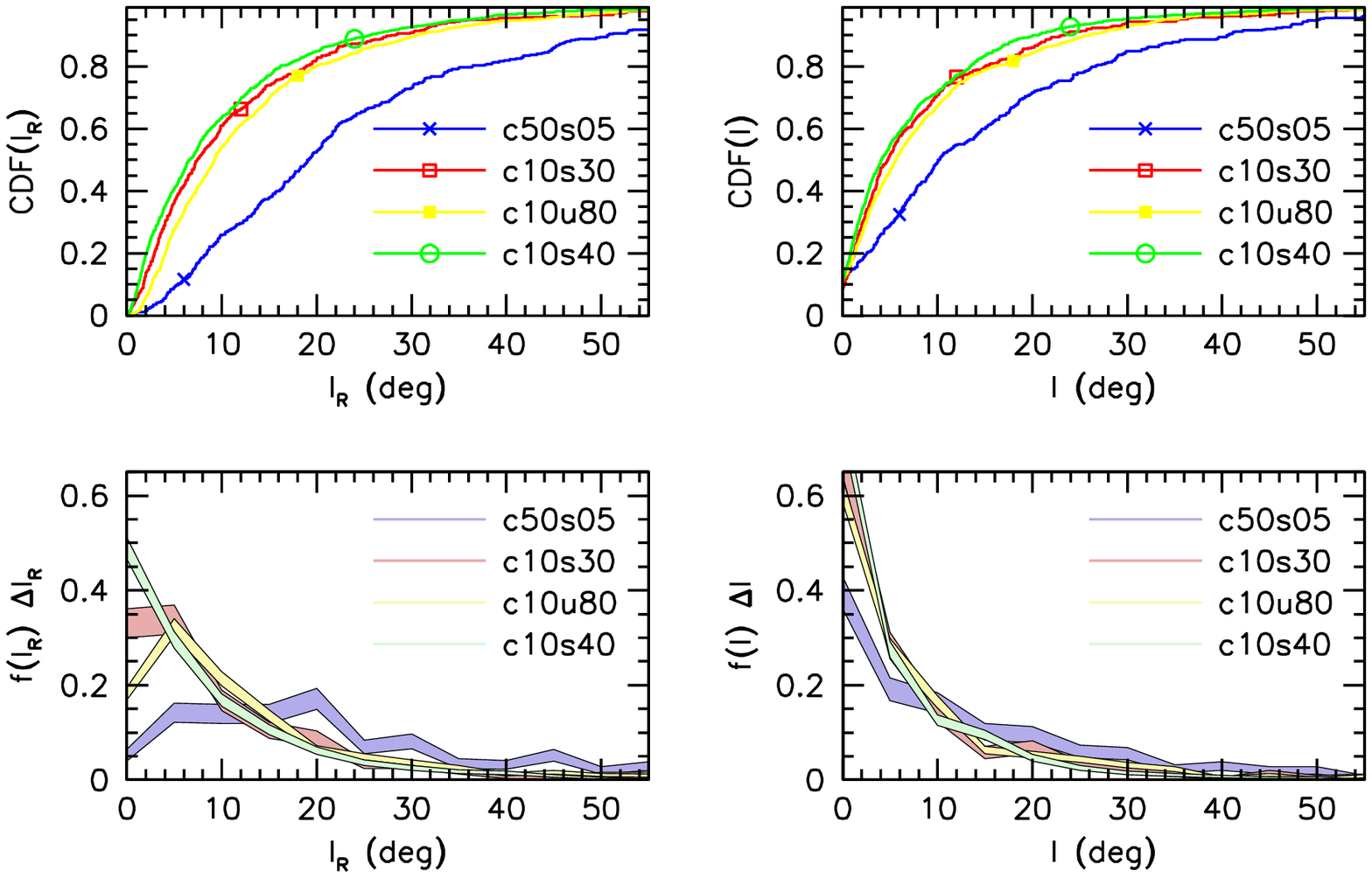}
\caption{ Cumulative (top row) and differential (bottom row) inclination
distributions of active systems, calculated with respect to the reference plane
of the initial conditions ($I_R$, left panels) and the invariable plane at the
end of the simulation ($I$, right panels). The bin size of the differential
distribution is $\Delta{}I_R=\Delta{}I=5^\circ$.
\label{incl_fig}}
\end{figure\biggie}

\begin{figure}[!ht]
\scl{.55}
\plotone{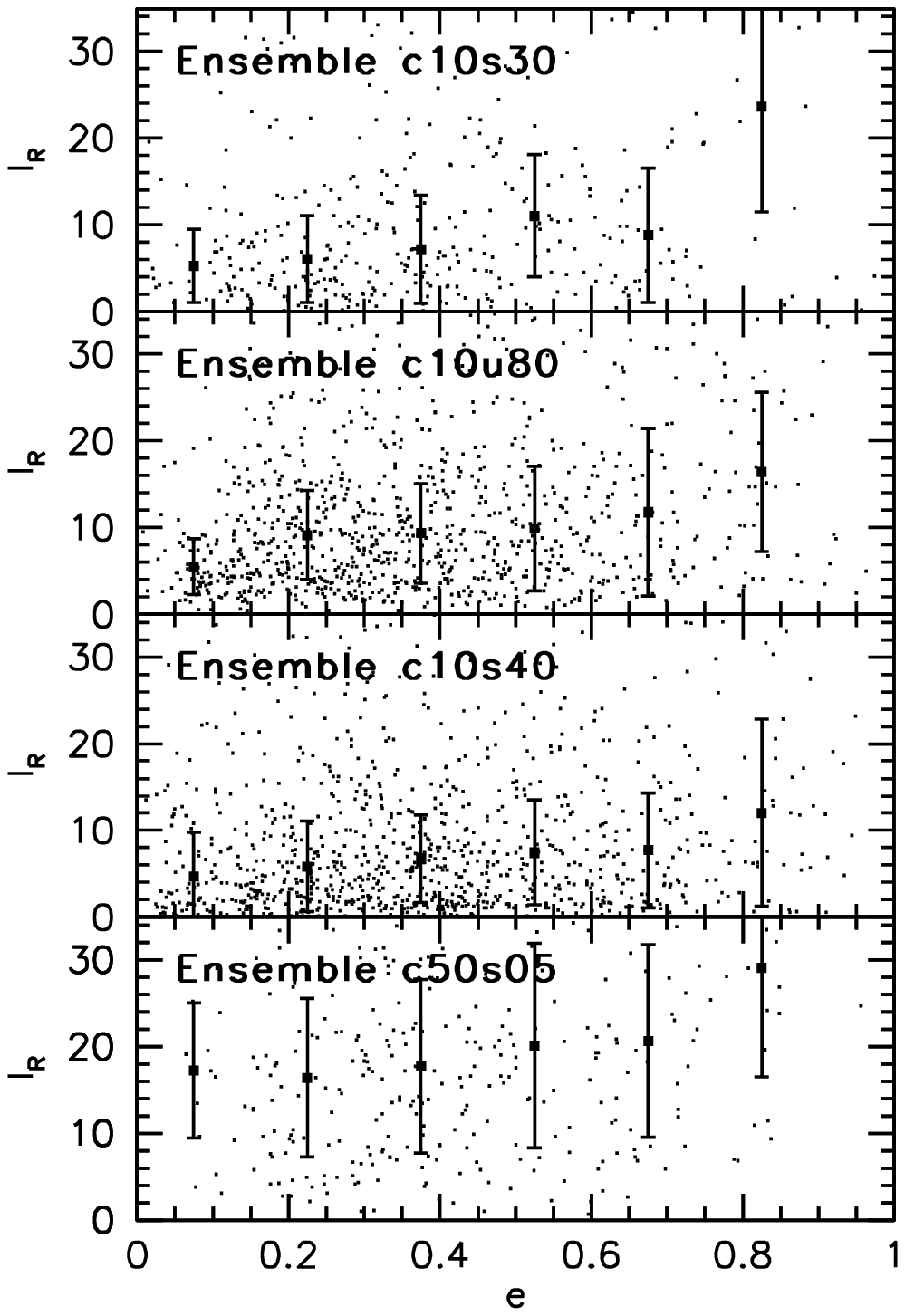}
\scl{1}
\caption{ Correlation of inclination and eccentricity in active systems. The
  inclinations are with respect to the reference plane of the initial
  conditions. \label{ei_fig}
}
\end{figure}

\begin{figure}[!ht]
\scl{.55}
\plotone{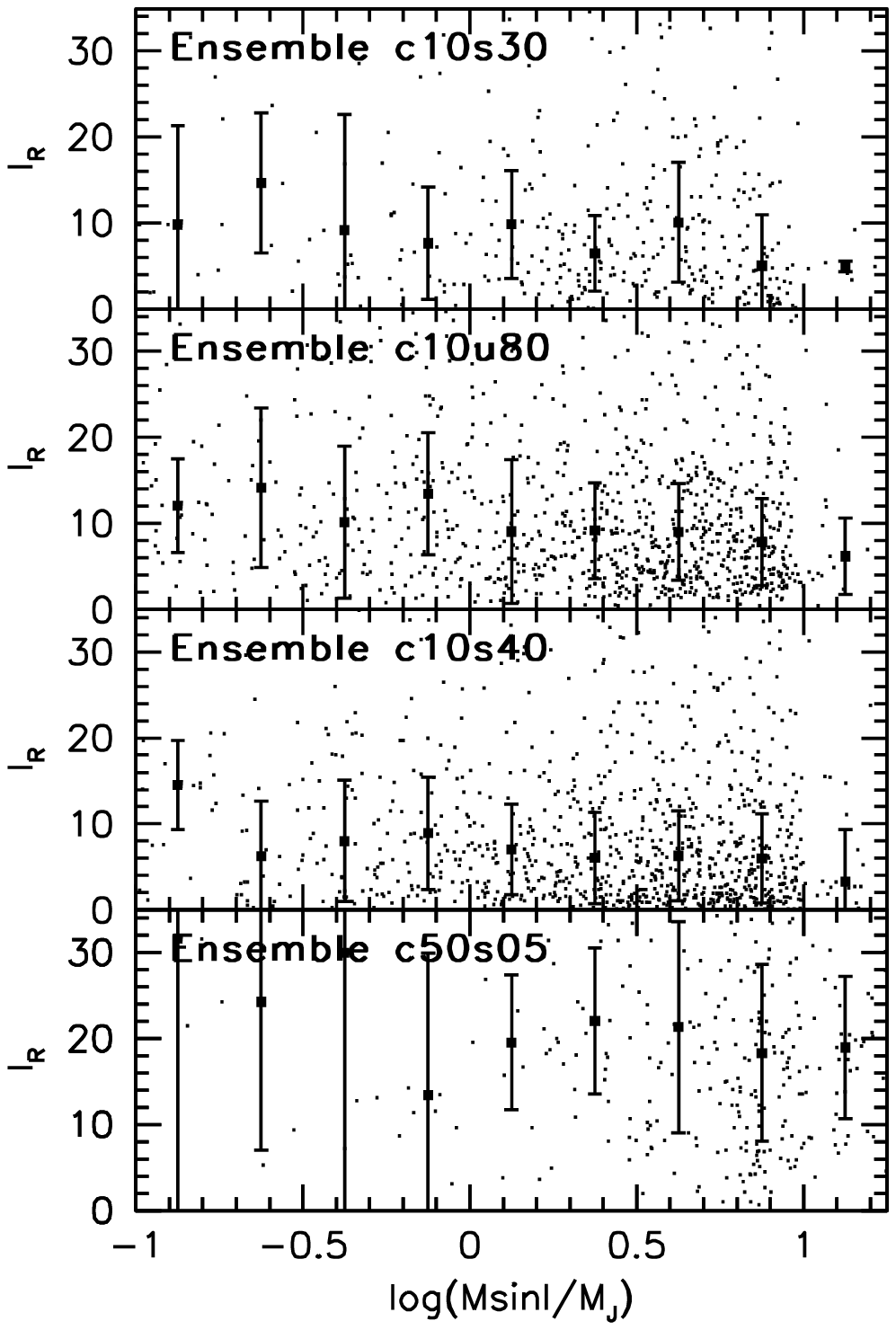}
\scl{1}
\caption{ Correlation of inclination and $M\sin I$ in active systems. The
  inclinations are with respect to the reference plane of the initial
  conditions.\label{logiplotComp}} 
\end{figure}

\begin{figure}[!ht]
\scl{.55}
\plotone{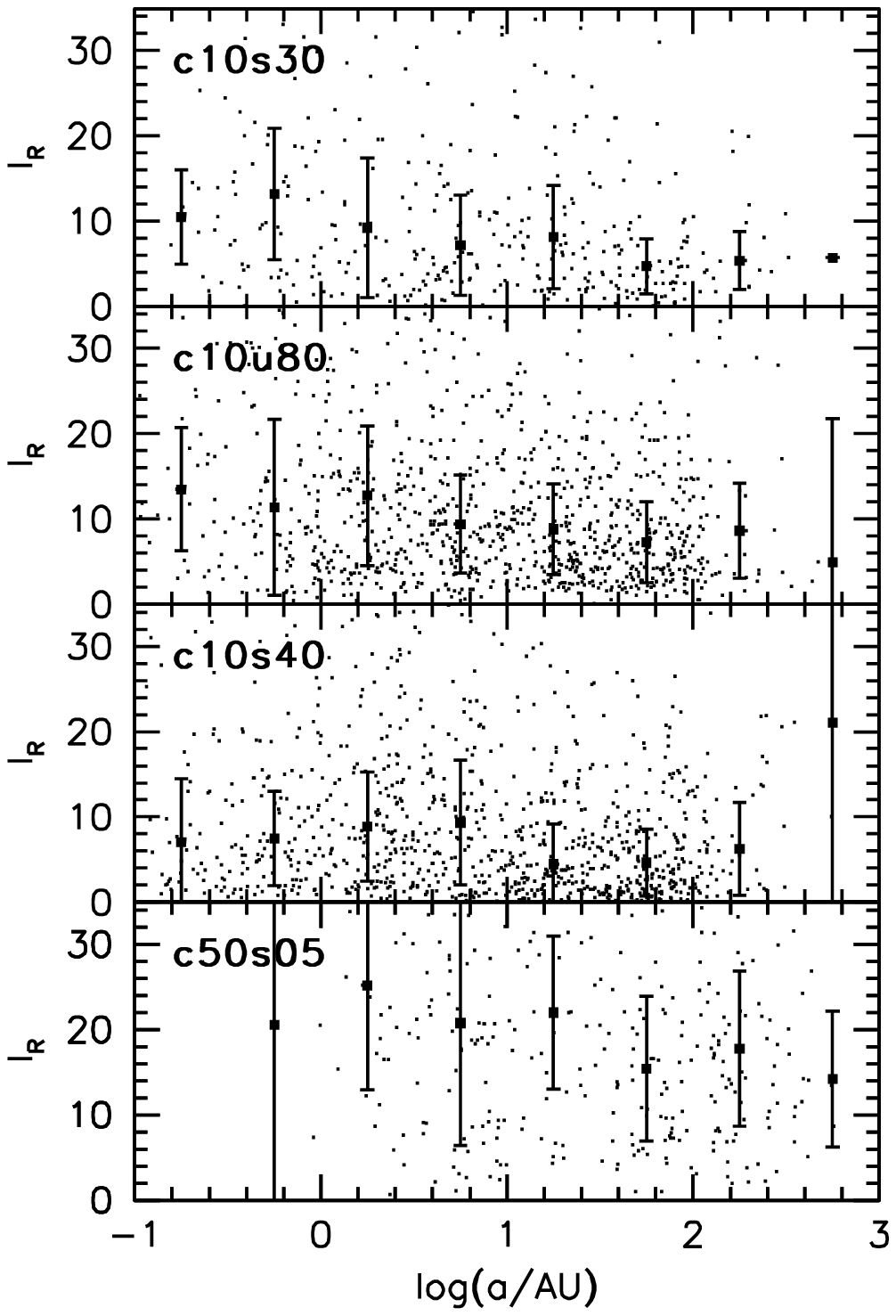}
\scl{1}
\caption{ Correlation of inclination and semimajor axis in active systems. The
  inclinations are with respect to the reference plane of the initial
  conditions.\label{logaiplotComp} 
}
\end{figure}

Cumulative and differential distributions of inclinations in active ensembles
are shown in Figure~\ref{incl_fig}. The left column shows the distribution of
inclinations relative to the symmetry plane of the initial conditions ($I_R$),
while on the right the inclinations are computed relative to the invariable
plane (the plane perpendicular to the total angular momentum) of each system
at the end of the simulation ($I$).

The inclinations $I$ referred to the invariable plane are in principle
measurable by precision astrometry of multi-planet systems. Such measurements
are currently out of reach of ground-based telescopes, but should become
feasible with the launch of the SIM PlanetQuest
mission\footnote{http://planetquest.jpl.nasa.gov/SIM/sim\_index.cfm}. On the
other hand, if the symmetry axis of the initial conditions is identified with
the axis of stellar rotation, the inclinations $I_R$ can be identified with
stellar spin-planetary orbit misalignments $\lambda$ and are measurable for
transiting planets using the Rossiter-McLaughlin effect \citep[the RM effect;
e.g.,][]{Ohta05,Winn05}. Other mechanisms, such as Kozai oscillations plus
tidal friction, can also cause spin-orbit misalignment \citep{2007ApJ...669.1298F}.

The final inclination distributions of the three active ensembles that start
with 10 planets appear similar, at least for
$I_R\gtrsim 4^\circ$, with medians $7^\circ < \widetilde{I}_R < 9^\circ$
($4^\circ < \widetilde{I} < 6^\circ$). For the 50 planet ensemble c50s05, the inclinations are
larger (median $\widetilde{I}_R = 19^\circ$ and $\widetilde{I} = 10^\circ$) and the
shape of the distribution is different.
All ensembles have a significant fraction of planets at high inclinations at
the end of the integrations; 10\% of planets of the c10- ensembles possess
inclinations $I_R \gtrsim 25^\circ$ ($I \gtrsim 20^\circ$), while 10\% of
c50s05 planets are inclined by $I_R > 51^\circ$ ($I > 40^\circ$). There is no
strong correlation of inclination and eccentricity (Figure~\ref{ei_fig})
except for the most eccentric planets ($e > 0.7$). A weak correlation exists
with mass (Figure~\ref{logiplotComp}), in the sense that the inclinations of
less massive planets are more easily excited than those of more massive ones.
Inclinations are weakly correlated with the semimajor axis 
(Figure~\ref{logaiplotComp}), in the sense that the inclinations of inner planets 
are on average higher than those of the outer. The strongest dependence is seen
for ensemble c10s30 ($d\widetilde{I}_R/d\log{a} = -2.8$~deg/dex), while the
effect is weakest for c10s40 ($d\widetilde{I}_R/d\log{a} = -0.9$~deg/dex).

Until recently, all measurements of the projected angle $\lambda$ between the
stellar spin axis and the planetary orbit axis from the RM effect were
 either small ($\lesssim 5^\circ$) or consistent with zero
\citep{Queloz00,Winn05,Winn06,Narita07,Wolf07}. While most of these involved
(possibly tidally evolved) hot Jupiters, no misalignment was found even in the
case of the significantly eccentric HAT-P-2b
\citep[$e=0.5$;][]{Bakos07,Winn07,Loeillet07}. A recent exception is HD 17156b
($P=21.2$~d, $e = 0.67$, \citealt{Fischer07,Barbieri07}), for which \cite{Narita07b} 
report a possible spin-orbit misalignment of
$\lambda = 62^\circ \pm 25^\circ$. At the nominal value $\lambda = 62^\circ$
this planet would be unusual in the context of our simulations
(a $1/100$ event in the c10-
ensembles and a 1/20 event in the c50s05 ensemble), even when the
inclination-semimajor axis correlation (Figure~\ref{logaiplotComp}) is taken into account.
However, if $\lambda$ is
lower by $1\sigma$ then the inclination of HD 17156b lies in more plausible
range ($p(I_R > 37^\circ) \gtrsim 0.05$ for the c10- ensembles, and $\sim 0.2$
for the c50s05 ensemble).

Collecting a larger sample of accurate measurements of the
projected spin-orbit angle $\lambda$ for eccentric planets may prove to be a
useful endeavor. The measurement of a significant misalignment in HD
17156b suggests that misalignments are common; more comprehensive simulations
of Stage 2 evolution can produce firm predictions for the dependence of the
final inclination distribution on the initial conditions (initial
number of planets, mass, semimajor axis, and inclination distribution, etc.);
and other processes such as Kozai oscillations yield equally firm but quite
different predictions.

\section{  A measure of dynamical activity   }
\label{hillpairs}

\begin{figure}
\scl{.7}
\plotone{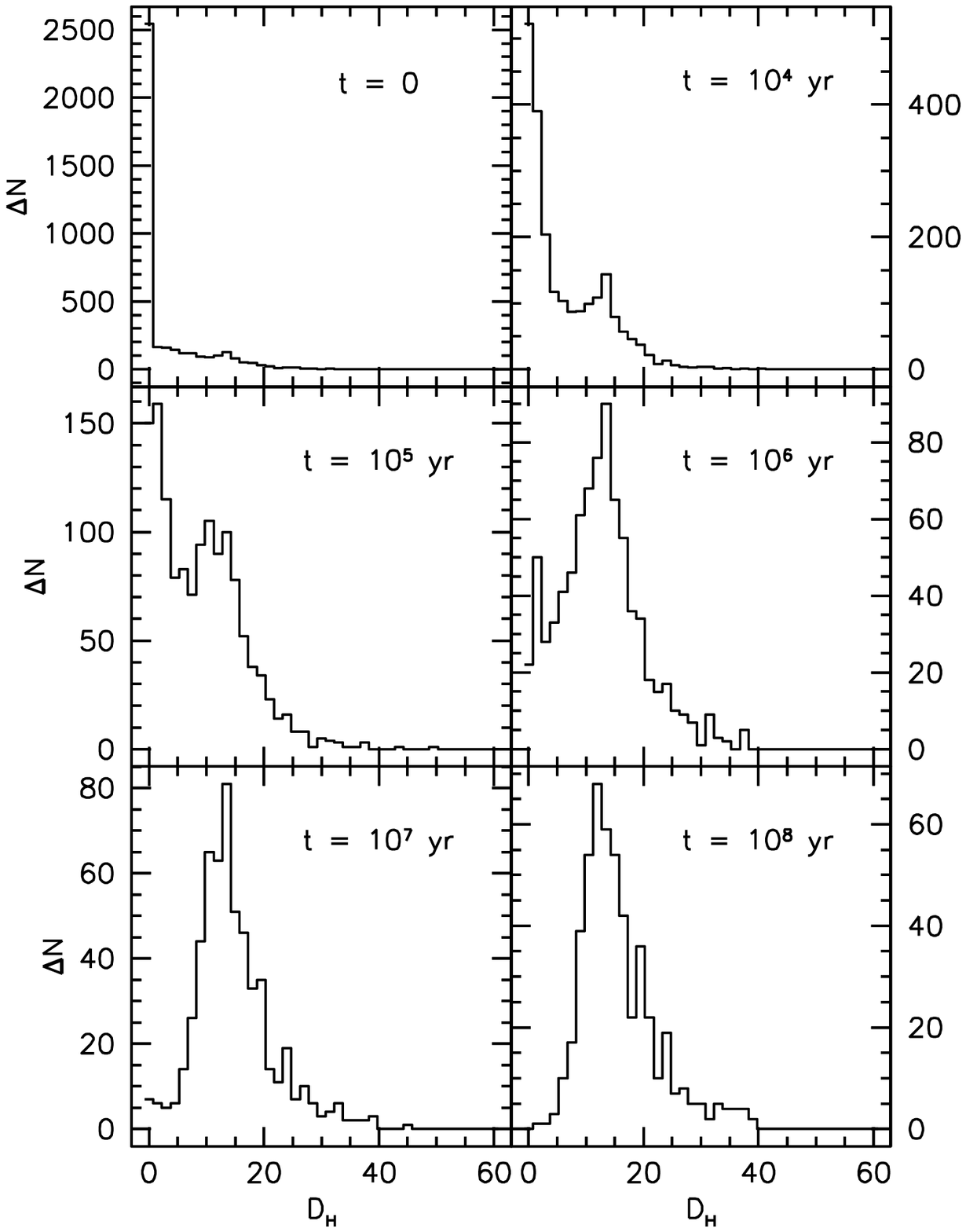}
\scl{1}
\caption{Evolution of the Hill neighbor separation distribution (NSD). The panels
show (left to right, top to bottom) the evolution of the NSD of ensemble
c10s40 at six different times.  Plotted are the number of planets in the
ensemble in bins of width $\Delta{}D_H=1.5$, where $D_H$ is the Hill
neighbor separation (\S\ref{hillpairs}).
\label{psdevol}}
\end{figure}

\begin{figure}
\scl{.6}
\plotone{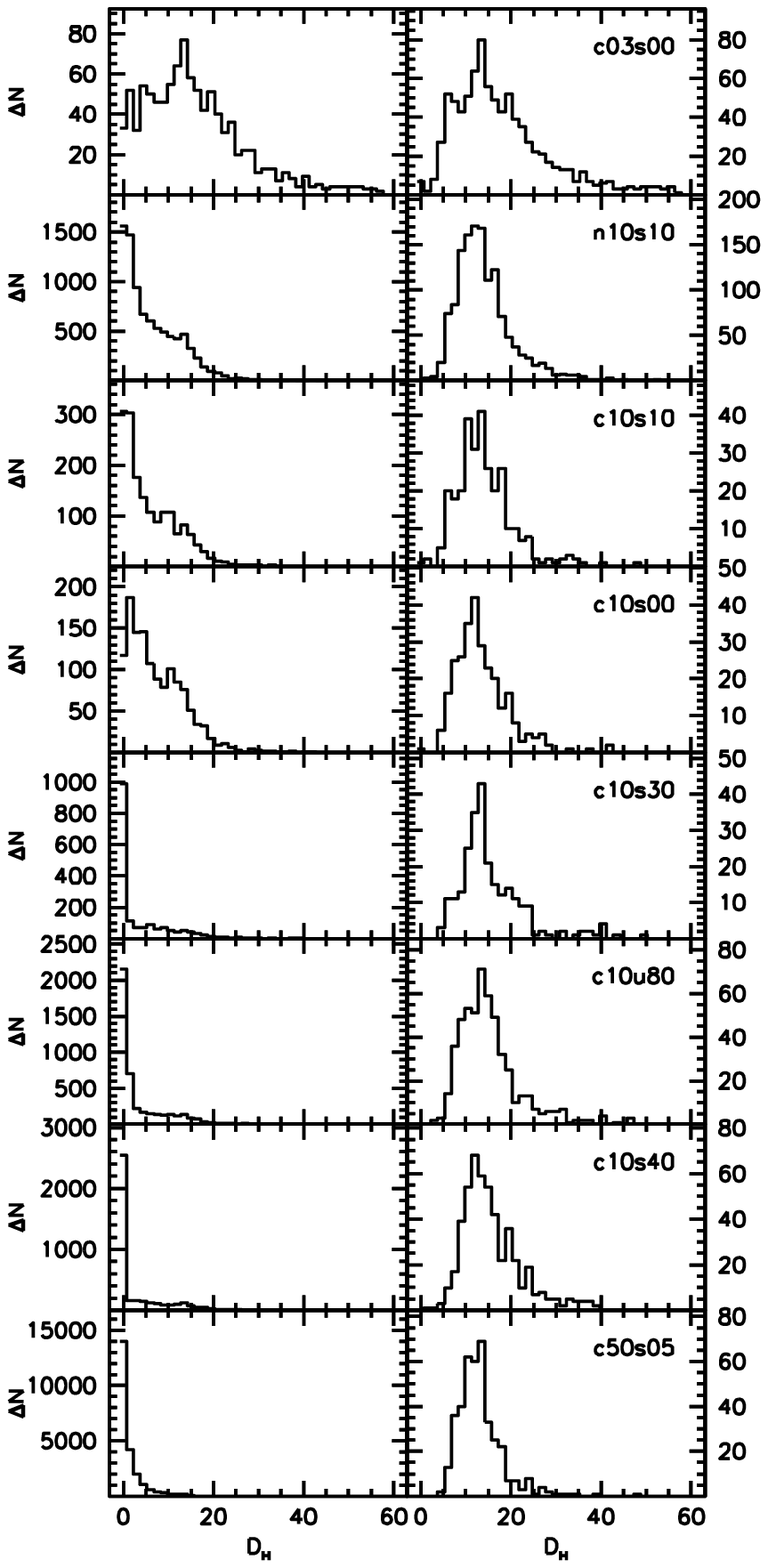}
\scl{1}
\caption{Comparison of initial and final Hill neighbor separation distributions
(NSDs) of simulated ensembles. Each row shows the initial ($t=0$, left panel)
and final ($t=10^8\yr$, right panel) NSD of the ensemble. Plotted are the
number of planets in the ensemble in each bin of width $\Delta{}D_H=1.5$.
\label{enspsdgrid}
}
\end{figure}

\begin{figure}
\plotone{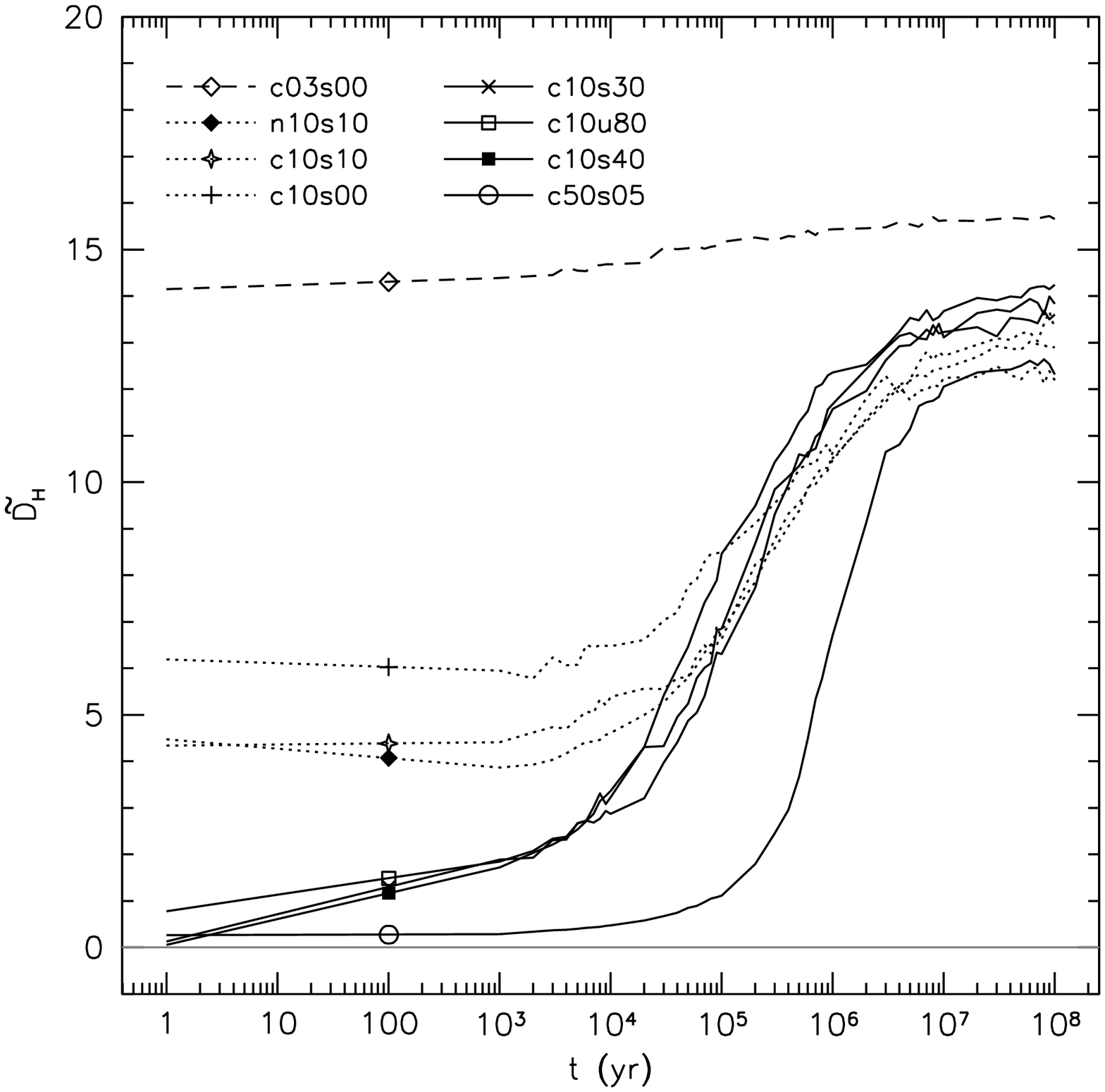}
\caption{Evolution of median Hill neighbor separation $\widetilde{D}_H$ with
time. The median is plotted against elapsed simulation time for each ensemble.
Active, partially active, and inactive ensembles are plotted with solid,
dotted and dashed lines, respectively.
\label{rhillevol}}
\end{figure}

The eccentricities of the three active ensembles with 10 initial planets, as
well as the $e > 0.2$ subsamples of all active ensembles,
are pairwise consistent with being drawn from the same underlying distribution.
The same holds true for the $e > 0.2$ subsamples of all partially active ensembles.
In \S\ref{eccdist} we have taken this
agreement as evidence that these ensembles have converged to the same
eccentricity distribution. We hypothesize that this distribution, empirically
described by a Schwarzschild distribution with $\sigma_e \sim 0.3$
(eq.~\ref{eq.sch}), is the equilibrium endpoint of ``dynamically active''
planetary systems, where by ``dynamically active'' we mean systems whose
planets experience strong mutual interactions and frequent encounters. We now
attempt to find an empirical measure of whether a planetary system with given
initial conditions will be dynamically active.

In the restricted three-body problem, the natural measure of the
radius of influence of a planet on a nearby test particle in a nearly
circular orbit is its Hill radius:
\begin{equation}
\label{eq.hillradius}
	R_h(r,M) = r \left[\frac{M}{3 M_\odot}\right]^{1/3}
\end{equation}
where $M$ and $r$ are the mass and orbital radius of the planet while
$M_\odot$ is the mass of the star.  In the case of two bodies with masses
$M_1$ and $M_2$ that are small compared to $M_\odot$, on nearly circular
orbits with similar radii $r$, the Hill radius is obtained from equation
(\ref{eq.hillradius}) by replacing $M$ with $M_1+M_2$
(\citealt{1986CeMec..38...67H}).  For the purposes of this paper, where we
must deal with planets having different masses and orbital radii, we define
the \emph{mutual Hill radius}
\begin{equation}
 R_{H} = \frac{1}{2} \left[ R_{h}(r_A,M_A) + R_{h}(r_B, M_B) \right]
\label{eq.mrhill}
\end{equation}
as the average of the Hill radii of the two individual planets. This
definition is somewhat arbitrary but reduces to the usual Hill radius when one
planet is much more massive than the other.

In the case of the general three-body problem \cite{1982CeMec..26..311M} and
\cite{Gladman93} have shown that two small planets on circular, nearly
coplanar orbits can have no close encounters (are ``Hill stable'') if their
semimajor axes are separated by\footnote{Note that \cite{Gladman93} defines the
mutual Hill radius as
\begin{equation}
	R_{H,G} = \frac{a_1 + a_2}{2} \left(\frac{m_1 + m_2}{3 M_\odot}\right)^{1/3}
\label{eq.mrhill.gladman}
\end{equation}
In the limit of equal-mass planets on nearby circular orbits, definitions
(\ref{eq.mrhill}) and (\ref{eq.mrhill.gladman}) differ by a factor of
$\sqrt[3]{2}$.} $a_2 - a_1 > 2\sqrt{3}
R_h[\half(a_1+a_2),M_1+M_2]$.
For systems of three or more planets, this criterion still approximately
predicts whether the system is unstable on short timescales. However, it does
not accurately predict long-term stability
(\citealt{1996Icar..119..261C}). Nevertheless, since it is usually the case
that only the one or two closest neighbors (expressed in Hill radii) are
responsible for most of the evolution, we can use the concept of Hill radii to
explore approximate criteria for whether a given planetary system is
dynamically active.

For a given planet $A$, we introduce the notion of its ``Hill neighbor'' $B$,
which is the planet of larger mass whose orbit comes closest, in terms of
mutual Hill radii, to the orbit of planet $A$. We define the \emph{Hill
neighbor separation}, $D_H$, to be the minimum distance between the orbital
ellipses of planet $A$ and its Hill neighbor $B$, expressed in mutual Hill
radii $R_H$. We use the minimum distance, instead of, say, the average
distance, because the mutual interaction of the two planets is strongest at
the point of closest approach.

For example, in a system with $N$ planets of different masses, there are $N-1$
Hill neighbors and $N-1$ Hill neighbor separations. In the n10s10 ensemble of
Table~\ref{tbl.ensembles}, there are 1000 systems of 10 planets each (at
$t=0$), and thus $9000$ Hill neighbors and $9000$ Hill neighbor separations.  We can
observe the time evolution of this ``Hill neighbor separation distribution'' (NSD)
and its statistical properties.

In Figure~\ref{psdevol} we show the NSD of ensemble c10s40 at $t=0$ (top left
panel) and its evolution from $t = 10^4\yr$ (top right panel) through $t =
10^8\yr$ (bottom right panel). At $t=0$ the systems of this ensemble are very
tightly packed, and will be unstable by virtually any criterion based on Hill
radii.  The ensemble reacts to this strongly unstable situation by removing
planets through collisions and ejections (see Figure~\ref{nvst}) and by
redistributing planets so as to increase the spacing between their orbits. As
a result, the number of planets with small $R_H$ decreases rapidly, and both
the peak and the median of the NSD shift towards higher values of $R_H$. At $t
= 10^8\yr$ all neighbors are separated by more that $4R_H$.

We repeat a similar analysis for all ensembles of
Table~\ref{tbl.ensembles}. Figure~\ref{enspsdgrid} shows the initial (left
column) and final (right column) NSD for each ensemble, starting with the one
``inactive'' ensemble in the top row, followed by three we classified as
``partially active'', and then the four ``active'' ensembles. A common
property of all active ensembles is the strong peak at small values of $D_H$
in the initial distributions. For example, while all four active ensembles
have initial median Hill neighbor separation $\widetilde{D}_H < 1$, the lowest
for a partially active ensemble (out of the three ensembles classified as
such) is $\widetilde{D}_H = 4.4$.

The final NSDs also share a number of common characteristics. All 
exhibit a sharp reduction in the number of objects at small values of
$D_H$. This gap near $D_H=0$ is more pronounced in active ensembles. All
examined NSDs peak at $D_H \simeq 12$, with the NSDs of active ensembles
(bottom four panels) having a similar unimodal distribution with a strong peak
at $D_H \simeq 12$, a width $\Delta{}D_H \sim 8$ (FWHM), and an
extended tail reaching to much larger values of $D_H$.

Figure~\ref{rhillevol} shows the evolution of the median Hill neighbor separation
$\widetilde{D}_H$ for the ensembles of Table~\ref{tbl.ensembles}. The division
into dashed, dotted and solid ensembles, corresponding to inactive, partially
active, and active ensembles, follows the convention adopted for
Figure~\ref{eccenCompInit} and the separation into top, middle and bottom
panels in Figure~\ref{eccenComp}.  The four ensembles with initial $\widetilde{D}_H <
1$ are all active; the three having $4 < \widetilde{D}_H < 7$ are partially
active, and the only ensemble with a large initial value
$\widetilde{D}_H\simeq 14$ is inactive. All active and partially active
ensembles converge to a final median Hill neighbor separation
$\widetilde{D}_H\simeq 12-14$ after $10^8\yr$.

We take these results to offer hope that the initial value of $\widetilde{D}_H$
may be used as a crude measure of the dynamical activity of an ensemble, and
hence as a predictor for the classification of the final eccentricity
distribution. A much more thorough exploration of
possible initial conditions is needed before we can tell whether this hope is
justified. In the meantime, our best guess is that systems with
$\widetilde{D}_H\lesssim 10$ are likely to be at least partially active. 

Other statistical measures may eventually prove to be more useful or reliable
in characterizing the dynamical activity of a planetary system. When devising
the one employed here, we were guided by the criteria that it should reflect
the level of short-term dynamical instability present in the system (satisfied
by expressing the distances to the closest more massive neighbor in Hill
radii), and that it should be applicable both to coplanar orbits and to orbits
with significant inclinations (requiring the relatively complex definition of
Hill neighbor separation). The statistic as defined above works reasonably well for
systems of the type existing in our simulations --- a few planets with a
limited range of masses (two decades). However, care must be taken when
applying it to (and interpreting it for) other types of systems, as we cannot prove
that it will work everywhere equally well, and strongly suspect that there
are possible, though perhaps pathological, planetary systems for which it does
not work at all.

\section{            Discussion                    }
\label{discussion}

As described in the Introduction, the possibility that planet-planet
interactions play a significant role in explaining the origin of the
extrasolar planet eccentricity distribution has been discussed since soon
after the first extrasolar planets were discovered. Most of these discussions
focused either on exploring the dynamics of simplified two or three planet
systems, or on following the dynamical evolution of planetary systems whose
initial conditions were inspired by planetesimal accretion theory\footnote{In
particular, the initial version of our paper was submitted to the arXiv
preprint server on the same day as papers by \cite{Chatterjee07} and
\cite{FR07}, who reach many of the same conclusions.}.

In this paper, we have integrated large ensembles of randomly constructed
planetary systems over 100 Myr timespans, simulating $PP\approx 5 \times
10^{12}$ planet-periods (eq.~\ref{eq:pp}). The output from these simulations, and future simulations
of this type, offers a rich resource for studies of Stage 2 planet evolution,
and here we have focused on only a few aspects of this evolution, in
particular the distributions of eccentricity, inclination, and separations.
Initially, we classify the ensembles according to their final eccentricity
distributions, and later show that this classification is strongly correlated
with the initial median Hill neighbor separation $\widetilde{D}_H$, in that all
ensembles classified as \emph{dynamically active} had $\widetilde{D}_H
\lesssim 1$ in the initial state. In all dynamically active ensembles that we
have examined, we obtain the same final eccentricity distribution for a
remarkably wide range of initial conditions. This distribution is described by
a Schwarzschild distribution (eq.~\ref{eq.sch}) with $\sigma_e \sim 0.3$. For
$e \gtrsim 0.2$ the final eccentricity distribution in our simulations of
active ensembles agrees with the observed eccentricity distribution of extrasolar
planets remarkably well. The excess of observed systems with $e\lesssim 0.2$
may reflect either a population of planetary systems that are not dynamically
active, or eccentricity damping by low-mass planets, planetesimals, or
residual gas. In the former case, comparison with
the observations suggests that about 25\% of the known extrasolar systems with
$P>20$~d are inactive, and 75\% active. 

We find little or no correlation of other parameters (semimajor axis,
planetary mass, and inclination) with the final eccentricity, although such
correlations are present during periods of dynamical instability early on in
the simulation.

This ``equilibrium'' distribution of eccentricity is a product of dynamical
relaxation of an initially unstable system. The distribution is mostly established
after $10^7\yr$ and remains stable to at least $10^8\yr$, where our
simulations end.  A few integrations have been carried to longer times, and
show no evidence of further evolution. By $10^8\yr$ most of the active
ensembles have only 2--3 remaining planets in the three decades of semimajor
axis that we originally populated. 

We find further that in all partially active and active ensembles that we
examined, $\widetilde{D}_H$ converges to a common value between $12$ and
$14$. In active ensembles, the distribution of $D_H$ converges to a common
shape as well, with a peak at $D_H \simeq 12$ and a width $\Delta{}D_H \simeq
8 $ (full width at half maximum). Thus \emph{both} the eccentricity
distribution and the distribution of Hill neighbor separations in active ensembles
appear to be a common endpoint of the dynamical relaxation process.

An aggressive interpretation of the similarity of the observed and theoretical
eccentricity distributions for $e\gtrsim 0.2$ is that the high eccentricities
of observed planets have arisen as an endpoint of dynamical relaxation, by
processes similar to those seen in the simulations of \S\ref{classification},
long after the initial stage of planet formation and dispersal of the
protoplanetary gas disk were complete.  This interpretation leads to a number
of interesting conclusions:

\begin{itemize}

\item There exists no single ``right'' ensemble of initial conditions at the
end of Stage 1. Instead, there is a multitude of substantially different
ensembles of initial conditions that lead to the same or similar final
outcomes, at least for the eccentricity distribution and the distribution of
Hill neighbor separations. An important corollary is that the details of initial
conditions are impossible to deduce from the ``relaxed'' component of the
observed eccentricity distribution, except to say that they are likely to be
in the ``active'' regime where dynamical evolution is strong enough to drive
the relaxation process.

\item In a large fraction of systems the final products of Stage 1 planet
formation must be dynamically active.  This is in principle possible for both
the planetesimal (e.g., \citealt{1998Icar..131..171K}) and gravitational
instability (e.g., \citealt{2000ApJ...536L.101B}) models. Of course, the
separation into Stage 1 and Stage 2 is somewhat artificial, since the initial
Stage 2 evolution occurs on timescales short compared to the likely duration
of Stage 1.

\item Planet-planet scattering in active ensembles changes the orbital planes
of planets, broadening the distribution of inclinations with respect to the
symmetry plane of the initial conditions. The same broadening is seen in
simulations of three-planet scattering by \cite{Chatterjee07}.  Assuming that
the host-star spin vector is parallel to the symmetry axis of the initial
protoplanetary disk, and that no other effects re-orient the spin axis of the
star or the invariable plane of the planets (e.g., \citealt{tre91}), this
inclination distribution is detectable, at least in principle, in transiting
planets through the Rossiter-McLaughlin effect. Given that the distribution of
misalignments depends on the initial conditions more strongly than the
distribution of eccentricities in active systems, its measurement in a large
sample of extrasolar planets may yield valuable information about the endpoint
of Stage 1 and the initial conditions for Stage 2.

\item The typical final number of giant planets in active systems is between
$2$ and $3$, at least over the range of three decades in semimajor axis that
we populate in the initial states (presumably other planets could form well
outside this semimajor axis range, and their interactions with our simulated
planets would be negligible). At $t=10^8\yr$, on average, 20\% of active
systems remain with only 1 planet, 75\% have two or three planets, and only
5\% have four or more planets, suggesting that most extrasolar planetary
systems should have 2--3 giant planets in this semimajor axis range (about one
giant planet per decade).  Consequently, extrasolar systems with a single
detected eccentric planet are likely to harbor at least one more planet of
comparable mass.  Observational data show long-term radial velocity trends
indicative of the presence of another planet in $\sim 50\%$ of the currently
known exoplanet systems (\citealt{Butler06}).  However, an equally interesting
prediction is that such systems are also \emph{not} likely to
harbor \emph{more} than one or two additional giant planets.

\item These predictions are consistent with current
observational data on the fraction of multi-planet systems. To compare our
simulations directly to the observations, we cull them at age $10^8\yr$ to
keep only those planets that produce a radial velocity semi-amplitude
$K>10\hbox{m s}^{-1}$ and have periods in the range 20~d$ < P< 2500$~d (the
lower limit eliminates hot Jupiters, and the upper limit approximates the
longest detectable orbital period). The culled ensembles c10s30, c10s40 and
c10u80 predict a ratio of single- to multi-planet systems $\simeq 86:14$
($84:16$ lowest, $89:11$ highest), in
excellent agreement with the $87:13$ ratio seen when the observed sample
subjected to same culling.  The agreement is only weakly sensitive to the
exact choice of threshold $K$ or the limits imposed on $P$. The prediction is
less good for the ratio of semimajor axes of the outer ($a_2$) and inner
($a_1$) planet in multi-planet systems, where the simulations typically peak 
at $4 < a_2/a_1 < 8$, while 80\% of the observed ratios are in $1 < a_2/a_1 < 4$
range. The distribution of semimajor axis ratios is not universal across the three
ensembles either, pointing to a dependence on initial conditions that warrants
further investigation.
Finally, the c50s05 ensemble, due to the efficient clearing of the zone with
$a < 1\au$ (Figure~\ref{lahist}), predicts no observed multi-planet systems at
all.

\item The typical final separation of planets in active multi-planet systems
should be $\widetilde{D}_H \simeq 12$--$14$. The determination of this
statistic in the currently known multiplanet systems is made difficult by the
unknown inclinations, both the distribution of inclinations relative to the
invariable plane and the inclination of the invariable plane to the line of
sight. We define a new statistic $\widetilde{D}_H'$, which is obtained from
$\widetilde{D}_H$ by replacing all planet masses $M$ by $M\sin I$, and compute
$\widetilde{D}_H'$ for our ensembles by assuming that the normal to the
invariable plane is distributed randomly and uniformly on a sphere and culling
the ensembles using the same criteria on period and velocity amplitude
described in the preceding paragraph. We obtain $\widetilde{D}_H'\simeq
12$--$13$. To compute the analogous statistic for the observations, we assume
that the inclinations relative to the invariable plane are described by a
Schwarzschild distribution (Eq.~\ref{eq.sch}) with $\sigma_I=10^\circ$, the
distribution of the nodal longitudes $\Omega$ is uniform, and we take the longitudes of periastron
$\varpi$ from \cite{Butler06}. We find $\widetilde{D}_H'\simeq 8$ for the 13 known multi-planet
systems. This agreement is probably adequate given the large statistical
errors, although it is also possible that the observed systems can be stable
at smaller values of  $\widetilde{D}_H'$ than our simulations because some of
the planets are stabilized by mean-motion resonances. 

\item The median separation $\widetilde{D}_H$ in the solar system is 15.6
(four giant planets) or 21.2 (all eight planets). Our results therefore
suggest that the solar system is inactive, which is consistent with the
observation that the eccentricities of planets in the solar system are much
lower than in extrasolar planetary systems\footnote{Although the solar system
appears to be inactive, in the sense that $\widetilde{D}_H$ appears to be
larger than needed for long-term stability of the existing planets, there are
almost no locations in the outer solar system (between Jupiter and Neptune) in
which additional planets could be inserted on stable orbits
\citep{hol97}.}. Our crude estimate in
\S\ref{comparisons} suggests that at least 25\% of extrasolar planetary 
systems are inactive, so in this respect the solar system is not unusual.

\item All active systems that we have examined eject a significant portion of
the initial mass in planets. On average, the active systems we simulated
ejected 50\% of the initial mass ($10 M_J$ ejected) if they started with 10
planets, and 80\% ($90 M_J$ ejected) if they started with 50. Therefore,
free-floating planets should be common and have a number density roughly
comparable to the number density of extrasolar planets. Such planets may be
detected by future microlensing surveys, or in open clusters as planetary-mass 
objects not bound to stars \citep{ZapateroOsorio00}.

\end{itemize} 

The constraints and limitations of the above conclusions have to be kept in
mind. The distribution of eccentricities is only a one-dimensional projection
of the multi-dimensional distribution of orbital elements, masses and other
properties, and it is likely that other statistics of this distribution do not
approach universal values in active Stage 2 evolution (for example, the mass
and semimajor axis distributions). It is also \emph{a priori} possible that
the equilibrium eccentricity distribution may be different for some ranges of
initial conditions that we failed to explore in this paper. A much broader
exploration of the possible initial conditions for Stage 2, and of the
distribution of orbital elements and masses at the end of Stage 2 evolution,
is needed to investigate these questions. It is particularly important to
explore (i) a steeper initial mass function (more low-mass and fewer high-mass
planets), since the final mass distributions in our simulations probably have
too few low-mass planets (Figure \ref{lmhist}); (ii) a minimum-mass cutoff
lower than our current value of $0.1M_J$, since a larger population of
low-mass planets may damp eccentricities; (iii) tidal dissipation from the
host star, which may affect orbits with pericenters less than a few stellar
radii; (iv) active systems that initially contain
giant planets only beyond a few AU, to investigate what fraction of giant
planets could acquire small semimajor axes through planet-planet
interactions; (v) active systems containing terrestrial planets. Although
investigating the wide range of possible initial conditions and final states
of Stage 2 evolution is a massive task, the only major resource required for
this investigation is processing time on cluster computers.

\acknowledgments

This research was supported in part by NASA grants NNG04H44g and NNX08AH83G,
and used computational facilities supported by NSF grant AST-0216105. MJ
gratefully acknowledges support from the Taplin Fellowship. The authors would
like to thank Bruce T. Draine and Robert Lupton for helpful discussions of 
computational and statistical aspects of this work; Peter Goldreich, Fred Rasio and Vicky
Kalogera for discussions; and the anonymous referee 
for thoughtful and detailed comments.

\bibliography{all}

\begin{thebibliography}{46}
\expandafter\ifx\csname natexlab\endcsname\relax\def\natexlab#1{#1}\fi

\bibitem[{{Adams} \& {Laughlin}(2003)}]{2003Icar..163..290A}
{Adams}, F.~C., \& {Laughlin}, G. 2003, Icarus, 163, 290

\bibitem[{{Bakos} {et~al.}(2007){Bakos}, {Kov{\'a}cs}, {Torres}, {Fischer},
  {Latham}, {Noyes}, {Sasselov}, {Mazeh}, {Shporer}, {Butler}, {Stefanik},
  {Fern{\'a}ndez}, {Sozzetti}, {P{\'a}l}, {Johnson}, {Marcy}, {Winn}, {Sip{\H
  o}cz}, {L{\'a}z{\'a}r}, {Papp}, \& {S{\'a}ri}}]{Bakos07}
{Bakos}, G.~{\'A}., {Kov{\'a}cs}, G., {Torres}, G., {Fischer}, D.~A., {Latham},
  D.~W., {Noyes}, R.~W., {Sasselov}, D.~D., {Mazeh}, T., {Shporer}, A.,
  {Butler}, R.~P., {Stefanik}, R.~P., {Fern{\'a}ndez}, J.~M., {Sozzetti}, A.,
  {P{\'a}l}, A., {Johnson}, J., {Marcy}, G.~W., {Winn}, J.~N., {Sip{\H o}cz},
  B., {L{\'a}z{\'a}r}, J., {Papp}, I., \& {S{\'a}ri}, P. 2007, \apj, 670, 826

\bibitem[{{Barbieri} {et~al.}(2007){Barbieri}, {Alonso}, {Laughlin},
  {Almenara}, {Bissinger}, {Davies}, {Gasparri}, {Guido}, {Lopresti},
  {Manzini}, \& {Sostero}}]{Barbieri07}
{Barbieri}, M., {Alonso}, R., {Laughlin}, G., {Almenara}, J.~M., {Bissinger},
  R., {Davies}, D., {Gasparri}, D., {Guido}, E., {Lopresti}, C., {Manzini}, F.,
  \& {Sostero}, G. 2007, \aap, 476, L13

\bibitem[{{Binney} \& {Tremaine}(2008)}]{1987gady.book.....B}
{Binney}, J., \& {Tremaine}, S. 2008, {Galactic Dynamics, 2nd ed.} (Princeton,
  NJ, Princeton University Press, 2008, 885 p.)

\bibitem[{{Boss}(2000)}]{2000ApJ...536L.101B}
{Boss}, A.~P. 2000, \apjl, 536, L101

\bibitem[{{Butler} {et~al.}(2006){Butler}, {Wright}, {Marcy}, {Fischer},
  {Vogt}, {Tinney}, {Jones}, {Carter}, {Johnson}, {McCarthy}, \&
  {Penny}}]{Butler06}
{Butler}, R.~P., {Wright}, J.~T., {Marcy}, G.~W., {Fischer}, D.~A., {Vogt},
  S.~S., {Tinney}, C.~G., {Jones}, H.~R.~A., {Carter}, B.~D., {Johnson}, J.~A.,
  {McCarthy}, C., \& {Penny}, A.~J. 2006, \apj, 646, 505

\bibitem[{{Chambers}(1999)}]{1999MNRAS.304..793C}
{Chambers}, J.~E. 1999, \mnras, 304, 793

\bibitem[{{Chambers} {et~al.}(1996){Chambers}, {Wetherill}, \&
  {Boss}}]{1996Icar..119..261C}
{Chambers}, J.~E., {Wetherill}, G.~W., \& {Boss}, A.~P. 1996, Icarus, 119, 261

\bibitem[{{Chatterjee} {et~al.}(2007){Chatterjee}, {Ford}, \&
  {Rasio}}]{Chatterjee07}
{Chatterjee}, S., {Ford}, E.~B., \& {Rasio}, F.~A. 2007, ArXiv Astrophysics
  e-prints, astro-ph/0703166

\bibitem[{{Cumming}(2004)}]{Cumming04}
{Cumming}, A. 2004, \mnras, 354, 1165

\bibitem[{{Fabrycky} \& {Tremaine}(2007)}]{2007ApJ...669.1298F}
{Fabrycky}, D., \& {Tremaine}, S. 2007, \apj, 669, 1298

\bibitem[{{Fischer} {et~al.}(2007){Fischer}, {Vogt}, {Marcy}, {Butler}, {Sato},
  {Henry}, {Robinson}, {Laughlin}, {Ida}, {Toyota}, {Omiya}, {Driscoll},
  {Takeda}, {Wright}, \& {Johnson}}]{Fischer07}
{Fischer}, D.~A., {Vogt}, S.~S., {Marcy}, G.~W., {Butler}, R.~P., {Sato}, B.,
  {Henry}, G.~W., {Robinson}, S., {Laughlin}, G., {Ida}, S., {Toyota}, E.,
  {Omiya}, M., {Driscoll}, P., {Takeda}, G., {Wright}, J.~T., \& {Johnson},
  J.~A. 2007, \apj, 669, 1336

\bibitem[{{Ford} {et~al.}(2001){Ford}, {Havlickova}, \& {Rasio}}]{FHR01}
{Ford}, E.~B., {Havlickova}, M., \& {Rasio}, F.~A. 2001, Icarus, 150, 303

\bibitem[{{Ford} \& {Rasio}(2007)}]{FR07}
{Ford}, E.~B., \& {Rasio}, F.~A. 2007, ArXiv Astrophysics e-prints, astro-ph/0703163

\bibitem[{{Ford} {et~al.}(2003){Ford}, {Rasio}, \& {Yu}}]{Ford03}
{Ford}, E.~B., {Rasio}, F.~A., \& {Yu}, K. 2003, in Astronomical Society of the
  Pacific Conference Series, Vol. 294, Scientific Frontiers in Research on
  Extrasolar Planets, ed. D.~{Deming} \& S.~{Seager}, 181--188

\bibitem[{{Gladman}(1993)}]{Gladman93}
{Gladman}, B. 1993, Icarus, 106, 247

\bibitem[{{Goldreich} \& {Sari}(2003)}]{GS03}
{Goldreich}, P., \& {Sari}, R. 2003, \apj, 585, 1024

\bibitem[{{Henon} \& {Petit}(1986)}]{1986CeMec..38...67H}
{Henon}, M., \& {Petit}, J.-M. 1986, Celestial Mechanics, 38, 67

\bibitem[{{Holman}(1997)}]{hol97}
{Holman}, M.~J. 1997, \nat, 387, 785

\bibitem[{{Kokubo} \& {Ida}(1998)}]{1998Icar..131..171K}
{Kokubo}, E., \& {Ida}, S. 1998, Icarus, 131, 171

\bibitem[{{Levison} {et~al.}(1998){Levison}, {Lissauer}, \&
  {Duncan}}]{1998AJ....116.1998L}
{Levison}, H.~F., {Lissauer}, J.~J., \& {Duncan}, M.~J. 1998, \aj, 116, 1998

\bibitem[{{Lin} \& {Ida}(1997)}]{LI97}
{Lin}, D.~N.~C., \& {Ida}, S. 1997, \apj, 477, 781

\bibitem[{{Loeillet} {et~al.}(2007){Loeillet}, {Shporer}, {Bouchy}, {Pont},
  {Mazeh}, {Beuzit}, {Boisse}, {Bonfils}, {Da Silva}, {Delfosse}, {Desort},
  {Ecuvillon}, {Forveille}, {Galland}, {Gallenne}, {Hebrard}, {Lagrange},
  {Lovis}, {Mayor}, {Moutou}, {Pepe}, {Perrier}, {Queloz}, {Segransan},
  {Sivan}, {Santos}, {Tsodikovich}, {Udry}, \& {Vidal-Madjar}}]{Loeillet07}
{Loeillet}, B., {Shporer}, A., {Bouchy}, F., {Pont}, F., {Mazeh}, T., {Beuzit},
  J.~L., {Boisse}, I., {Bonfils}, X., {Da Silva}, R., {Delfosse}, X., {Desort},
  M., {Ecuvillon}, A., {Forveille}, T., {Galland}, F., {Gallenne}, A.,
  {Hebrard}, G., {Lagrange}, A.~M., {Lovis}, C., {Mayor}, M., {Moutou}, C.,
  {Pepe}, F., {Perrier}, C., {Queloz}, D., {Segransan}, D., {Sivan}, J.~P.,
  {Santos}, N.~C., {Tsodikovich}, Y., {Udry}, S., \& {Vidal-Madjar}, A. 2007,
  ArXiv e-prints, ArXiv:0707.0679

\bibitem[{{Marchal} \& {Bozis}(1982)}]{1982CeMec..26..311M}
{Marchal}, C., \& {Bozis}, G. 1982, Celestial Mechanics, 26, 311

\bibitem[{{Marcy} {et~al.}(2005){Marcy}, {Butler}, {Fischer}, {Vogt}, {Wright},
  {Tinney}, \& {Jones}}]{2005PThPS.158...24M}
{Marcy}, G., {Butler}, R.~P., {Fischer}, D., {Vogt}, S., {Wright}, J.~T.,
  {Tinney}, C.~G., \& {Jones}, H.~R.~A. 2005, Progress of Theoretical Physics
  Supplement, 158, 24

\bibitem[{{Marzari} \& {Weidenschilling}(2002)}]{MW02}
{Marzari}, F., \& {Weidenschilling}, S.~J. 2002, Icarus, 156, 570

\bibitem[{{Naef} {et~al.}(2001){Naef}, {Latham}, {Mayor}, {Mazeh}, {Beuzit},
  {Drukier}, {Perrier-Bellet}, {Queloz}, {Sivan}, {Torres}, {Udry}, \&
  {Zucker}}]{2001A&A...375L..27N}
{Naef}, D., {Latham}, D.~W., {Mayor}, M., {Mazeh}, T., {Beuzit}, J.~L.,
  {Drukier}, G.~A., {Perrier-Bellet}, C., {Queloz}, D., {Sivan}, J.~P.,
  {Torres}, G., {Udry}, S., \& {Zucker}, S. 2001, \aap, 375, L27

\bibitem[{{Nagasawa} {et~al.}(2008){Nagasawa}, {Ida}, \& {Bessho}}]{Nagasawa08}
{Nagasawa}, M., {Ida}, S., \& {Bessho}, T. 2008, \apj, 678, 498

\bibitem[{{Narita} {et~al.}(2007{\natexlab{a}}){Narita}, {Enya}, {Sato},
  {Ohta}, {Winn}, {Suto}, {Taruya}, {Turner}, {Aoki}, {Yoshii}, {Yamada}, \&
  {Tamura}}]{Narita07}
{Narita}, N., {Enya}, K., {Sato}, B., {Ohta}, Y., {Winn}, J.~N., {Suto}, Y.,
  {Taruya}, A., {Turner}, E.~L., {Aoki}, W., {Yoshii}, M., {Yamada}, T., \&
  {Tamura}, Y. 2007{\natexlab{a}}, \pasj, 59, 763

\bibitem[{{Narita} {et~al.}(2007{\natexlab{b}}){Narita}, {Sato}, {Ohshima}, \&
  {Winn}}]{Narita07b}
{Narita}, N., {Sato}, B., {Ohshima}, O., \& {Winn}, J.~N. 2007{\natexlab{b}},
  ArXiv e-prints, ArXiv:0712.2569

\bibitem[{{Navarro} {et~al.}(1997){Navarro}, {Frenk}, \&
  {White}}]{1997ApJ...490..493N}
{Navarro}, J.~F., {Frenk}, C.~S., \& {White}, S.~D.~M. 1997, \apj, 490, 493

\bibitem[{{Ohta} {et~al.}(2005){Ohta}, {Taruya}, \& {Suto}}]{Ohta05}
{Ohta}, Y., {Taruya}, A., \& {Suto}, Y. 2005, \apj, 622, 1118

\bibitem[{{Papaloizou} \& {Terquem}(2001)}]{2001MNRAS.325..221P}
{Papaloizou}, J.~C.~B., \& {Terquem}, C. 2001, \mnras, 325, 221

\bibitem[{{Press} {et~al.}(1992){Press}, {Teukolsky}, {Vetterling}, \&
  {Flannery}}]{1992nrca.book.....P}
{Press}, W.~H., {Teukolsky}, S.~A., {Vetterling}, W.~T., \& {Flannery}, B.~P.
  1992, {Numerical recipes in C. The art of scientific computing} (Cambridge
  University Press, 2nd ed.)

\bibitem[{{Queloz} {et~al.}(2000){Queloz}, {Eggenberger}, {Mayor}, {Perrier},
  {Beuzit}, {Naef}, {Sivan}, \& {Udry}}]{Queloz00}
{Queloz}, D., {Eggenberger}, A., {Mayor}, M., {Perrier}, C., {Beuzit}, J.~L.,
  {Naef}, D., {Sivan}, J.~P., \& {Udry}, S. 2000, \aap, 359, L13

\bibitem[{{Rasio} \& {Ford}(1996)}]{RF96}
{Rasio}, F.~A., \& {Ford}, E.~B. 1996, Science, 274, 954

\bibitem[{{Rauch} \& {Holman}(1999)}]{1999AJ....117.1087R}
{Rauch}, K.~P., \& {Holman}, M. 1999, \aj, 117, 1087

\bibitem[{{Tabachnik} \& {Tremaine}(2002)}]{2002MNRAS.335..151T}
{Tabachnik}, S., \& {Tremaine}, S. 2002, \mnras, 335, 151

\bibitem[{{Tremaine}(1991)}]{tre91}
{Tremaine}, S. 1991, Icarus, 89, 85

\bibitem[{{Weidenschilling} \& {Marzari}(1996)}]{WM96}
{Weidenschilling}, S.~J., \& {Marzari}, F. 1996, \nat, 384, 619

\bibitem[{{Winn} {et~al.}(2006){Winn}, {Johnson}, {Marcy}, {Butler}, {Vogt},
  {Henry}, {Roussanova}, {Holman}, {Enya}, {Narita}, {Suto}, \&
  {Turner}}]{Winn06}
{Winn}, J.~N., {Johnson}, J.~A., {Marcy}, G.~W., {Butler}, R.~P., {Vogt},
  S.~S., {Henry}, G.~W., {Roussanova}, A., {Holman}, M.~J., {Enya}, K.,
  {Narita}, N., {Suto}, Y., \& {Turner}, E.~L. 2006, \apjl, 653, L69

\bibitem[{{Winn} {et~al.}(2007){Winn}, {Johnson}, {Peek}, {Marcy}, {Bakos},
  {Enya}, {Narita}, {Suto}, {Turner}, \& {Vogt}}]{Winn07}
{Winn}, J.~N., {Johnson}, J.~A., {Peek}, K.~M.~G., {Marcy}, G.~W., {Bakos},
  G.~{\'A}., {Enya}, K., {Narita}, N., {Suto}, Y., {Turner}, E.~L., \& {Vogt},
  S.~S. 2007, \apjl, 665, L167

\bibitem[{{Winn} {et~al.}(2005){Winn}, {Noyes}, {Holman}, {Charbonneau},
  {Ohta}, {Taruya}, {Suto}, {Narita}, {Turner}, {Johnson}, {Marcy}, {Butler},
  \& {Vogt}}]{Winn05}
{Winn}, J.~N., {Noyes}, R.~W., {Holman}, M.~J., {Charbonneau}, D., {Ohta}, Y.,
  {Taruya}, A., {Suto}, Y., {Narita}, N., {Turner}, E.~L., {Johnson}, J.~A.,
  {Marcy}, G.~W., {Butler}, R.~P., \& {Vogt}, S.~S. 2005, \apj, 631, 1215

\bibitem[{{Wisdom} \& {Holman}(1991)}]{1991AJ....102.1528W}
{Wisdom}, J., \& {Holman}, M. 1991, \aj, 102, 1528

\bibitem[{{Wolf} {et~al.}(2007){Wolf}, {Laughlin}, {Henry}, {Fischer}, {Marcy},
  {Butler}, \& {Vogt}}]{Wolf07}
{Wolf}, A.~S., {Laughlin}, G., {Henry}, G.~W., {Fischer}, D.~A., {Marcy}, G.,
  {Butler}, P., \& {Vogt}, S. 2007, \apj, 667, 549

\bibitem[{{Wu} \& {Murray}(2003)}]{wm03}
{Wu}, Y., \& {Murray}, N. 2003, \apj, 589, 605

\bibitem[{{Zapatero Osorio} {et~al.}(2000){Zapatero Osorio}, {B{\'e}jar},
  {Mart{\'{\i}}n}, {Rebolo}, {y Navascu{\'e}s}, {Bailer-Jones}, \&
  {Mundt}}]{ZapateroOsorio00}
{Zapatero Osorio}, M.~R., {B{\'e}jar}, V.~J.~S., {Mart{\'{\i}}n}, E.~L.,
  {Rebolo}, R., {y Navascu{\'e}s}, D.~B., {Bailer-Jones}, C.~A.~L., \& {Mundt},
  R. 2000, Science, 290, 103

\end{thebibliography}
\end{document}